\documentclass{stsci_report}

\usepackage{graphicx}
\usepackage{listings}
\usepackage{xcolor}
\usepackage{amssymb}

\usepackage{siunitx}
\usepackage{todonotes}
\usepackage{pdfpages}
\usepackage{hyperref}
\usepackage[parfill]{parskip}
\usepackage{dirtytalk}
\usepackage[section]{placeins}
\usepackage{makecell}
\usepackage{caption}
\usepackage{aas_macros}

\usepackage{float}
\usepackage{graphicx}
\usepackage{siunitx}
\usepackage{todonotes}
\usepackage{pdfpages}
\usepackage{tabularx}
\usepackage{multirow}
\usepackage{longtable}
\usepackage{caption}
\usepackage{subcaption}
\usepackage{aas_macros}
\usepackage[bottom]{footmisc}
\usepackage{wrapfig}
\usepackage{amsmath}
\usepackage{colortbl}
\usepackage{tabularray }
\usepackage{setspace} 
\usepackage{natbib}
\setcitestyle{aysep={}}
\usepackage{listings}
\usepackage{ragged2e}

\newcommand{\specialcell}[2][c]{%
  \begin{tabular}[#1]{@{}c@{}}#2\end{tabular}}

\DeclareCaptionType{equ}[][]

\usepackage{amsmath} 

\definecolor{codegreen}{rgb}{0,0.6,0}
\definecolor{codegray}{rgb}{0.5,0.5,0.5}
\definecolor{codepurple}{rgb}{0.58,0,0.82}
\definecolor{backcolour}{rgb}{0.95,0.95,0.92}

\lstdefinestyle{mystyle}{
  backgroundcolor=\color{backcolour},   
  commentstyle=\color{blue},
  keywordstyle=\color{codegreen},
  numberstyle=\tiny\color{codegray},
  stringstyle=\color{codepurple},
  basicstyle=\ttfamily\footnotesize,
  breakatwhitespace=false,         
  breaklines=true,                 
  captionpos=b,                    
  keepspaces=true,                 
  numbers=left,                    
  numbersep=5pt,                  
  showspaces=false,                
  showstringspaces=false,
  showtabs=false,                  
  tabsize=1
}

\makeatletter
\renewcommand\@makefntext[1]{%
  \raggedright\noindent\@makefnmark~#1}
\makeatother

\copyrighttext{Copyright\copyright\ \the\year\ The Association of Universities for Research in Astronomy, Inc. All Rights Reserved.}

\presubtitle{Instrument Science Report WFC3 2026-02}
\title{Updates to WFC3/UVIS Encircled Energy Values in Select Filters}
\author{Anne O'Connor, Jennifer Mack, Varun Bajaj}

\date{July 21, 2026}

\begin{document}

\maketitle
\abstract{We present updated encircled energy (EE) curves for a subset of UVIS filters. These are derived from a reanalysis of the drizzled Point Spread Function (PSF) data underlying the current EE calibration together with new measurements from deep observations of the PSF wings. Improved centroiding and analysis techniques applied to the drizzled PSFs produce more accurate EE values at small radii ($r \le 10$ pixels), bringing the results into closer agreement with a large archival PSF study \citep{2025wfc..rept....5H}. At large radii, we leverage deep observations of the PSF wings in six filters and compare the fraction of light between 2\unit{\arcsecond} and 6\unit{\arcsecond} with predictions from an optical model of the PSF \citep{2009wfc..rept...38H}.  
For filters with pivot wavelengths $ \gtrsim{4000} $ \AA, the model agrees with the empirical data, but over-estimates the EE for UV filters by $\sim0.5\%$ at 2\unit{\arcsecond}. 
The revised EE solutions affect the UVIS zeropoints, which are derived from aperture photometry in a 0.4\unit{\arcsecond} (10 pixel) radius and corrected to 6\unit{\arcsecond} using EE tables \citep{2022AJ....164...32C}. 
Overall, the impact is small; the EE at 0.4\unit{\arcsecond} is larger by $ \gtrsim{0.5}$\% (0.005 mag) for several filters (F218W, F225W, F275W, F775W, F814W, F845M), especially for UVIS2. 
In contrast, the EE value is generally smaller at 0.4\unit{\arcsecond} for long-pass filters (F200LP, F350LP, F850LP), with F850LP differing by $\sim 2 $\%  (0.02 mag).  
Updated EE tables will be delivered together with a revised set of UVIS inverse sensitivity tables and zeropoints later in 2026. In the interim, we provide EE tables for commonly used filters in Appendix A.
}

\section{Introduction}
\label{sec:Intro}
Encircled energy (EE) describes the fraction of a point source’s total flux enclosed within a circular aperture of radius \textit{r}. 
EE measurements are essential to the WFC3 photometric calibration, where aperture photometry of HST flux standards is computed using an 0.4\unit{\arcsecond} (10-pixel) radius aperture and then corrected using the EE tables to compute the total flux (in electrons per second) in an `infinite aperture'. Comparing with the known flux for each star, an inverse sensitivity value may be computed for each filter, representing the flux density of a source (erg cm\textsuperscript{-2} s\textsuperscript{-1} Å\textsuperscript{-1}) that produces a response of one electron per second in the bandpass (see the WFC3 Data Handbook Section 9.1 for more details). 

\subsection{EE History}
\label{sec:EE_history}

Following the alignment of WFC3 to the OTA (Optical Telescope Assembly), a series of observations were obtained to demonstrate the optical performance for the UVIS channel as a function of wavelength, as described by its point-spread function (PSF), i.e., the spatial distribution of the flux in an image of a point source. This was used to produce an optical model of the PSF, incorporating knowledge of the pupil geometry, residual aberration, mid-frequency wavefront error of the OTA, detector charge diffusion effects, and first-order geometric distortion (Marinelli et al. 2025, WFC3 Instrument Handbook).

The optical model was adjusted using in-flight exposures in two filters, F275W and F625W, and represents the mean EE over the field of view (averaged over five positions). EE tables provided the ratio of the enclosed flux (in electrons per second) at different aperture radii and the flux at infinity, defined at a radius of 6 arcseconds (6\unit{\arcsecond}  $ \approx 150$ pixels) for both WFC3 channels (UVIS and IR). Tabular values were provided for a grid of aperture radii from 0.1\unit{\arcsecond} to 2.0\unit{\arcsecond}, and the first UVIS photometric zeropoints \citep{2009wfc..rept...31K} used the model EE values to correct for the fraction of light between 10 pixels (0.4\unit{\arcsecond}) and 6.0\unit{\arcsecond}.  The model provides a grid of EE values between $2,000\AA$ and $10,000\AA$ in $1,000\AA$ increments. EE corrections were therefore linearly interpolated to the pivot wavelength for each filter. 

In 2016, repeated observations of CALSPEC\footnote{CALSPEC contains composite stellar spectra that are flux standards for HST photometric calibration. https://www.stsci.edu/hst/instrumentation/reference-data-for-calibration-and-tools/astronomical-catalogs/calspec} spectrophotometric standards in the corner subarrays of UVIS1 (C512A) and UVIS2 (C512C) were used to derive EE curves for all 42 UVIS filters at aperture radii from 0.04\unit{\arcsecond} - 6\unit{\arcsecond} (1 - 150 pixels). For most UVIS filters, the drizzled PSFs provided reliable measurements out to an aperture radius of $r \approx 35$ pixels (1.4\unit{\arcsecond}) \citep{2016wfc..rept....3D}. The EE curves were therefore normalized and spliced to the optical model EE by linearly interpolating between the 1.0\unit{\arcsecond} and 1.5\unit{\arcsecond} apertures provided in the table. 

In 2020, updated EE curves were computed for a subset of filters using drizzled PSFs derived by stacking a much larger set of monitoring observations obtained between 2009 and 2019, with an explicit correction for filter-dependent sensitivity losses on each CCD. The resulting PSFs had improved signal-to-noise compared to the 2016 PSFs, so the EE curves were no longer spliced to the optical model EE, but normalized directly to the observed flux at 6\unit{\arcsecond} \citep{2022AJ....164...32C, 2022wfc..rept....2M}. 

\subsection{Motivation for Revisiting the EE}
\label{sec:motivation}

A recent study of archival PSFs \citep{2025wfc..rept....5H} used focus-diverse empirical PSFs by \cite{2018wfc..rept...14A} in five UVIS filters to derive aperture corrections as a function of detector position and focus level. The authors used isolated stars in archival UVIS images to measure aperture corrections between 5 and 10 pixels and compare with the results derived from EE tables. For three filters (F336W, F438W, and F606W), based on 2016 EE curves, the aperture correction values are generally consistent with the EE tables for both UVIS1 and UVIS2. For the two filters based on 2020 EE curves (F275W and F814W), large discrepancies of $\sim0.02 - 0.03$ mag in the UVIS2 aperture correction results were found, larger than the $\sim0.01$ mag variation across all focus levels at a given detector position. This motivated a re-examination of the 2020 empirical PSFs used to derive the EE tables to attempt to reproduce the previous solutions.

\section{Methodology}
\label{sec:methodology}

In this report, we re-compute the EE for a subset of WFC3/UVIS filters using the drizzled PSFs by \cite{2022wfc..rept....2M} and the 2016 drizzled PSFs by \cite{2016wfc..rept....3D}, both of  which are used for the current UVIS photometric calibration by \cite{2022AJ....164...32C}.  Using updated methodology, we reanalyze the drizzled PSFs, referred to as ``short-stacks", since they were created from stacking many short exposures. Additonally, we leverage new ``deep PSF" calibration observations to measure the fraction of light in the PSF at large aperture radii. 

The updated methodology is made up of a few key steps. First,  we calculate a refined centroid for each star in the short-stack images by fitting a 2-D Gaussian profile to the data within a small cutout box around the initial centroid, ensuring sub-pixel accuracy. Using this improved centroid position, we compute the curve of growth with the \textit{Photutils} photometry package \citep{2016ascl.soft09011B}. Count rates are measured in circular apertures with radii from 1 to 150 pixels (6\unit{\arcsecond}).  The sky background is estimated for each image using the $3\sigma$-clipped mean within an annulus spanning 156–165 pixels, for consistency with the annulus used to derive the UVIS zeropoints \citep{2022AJ....164...32C}.

To determine which filters have adequate signal in the short-stack images, we visually examine the azimuthal average curves, treating the radius at which they become noisy as an indicator of insufficient signal to reliably measure the EE out to 6\unit{\arcsecond}. We compare the EE from short-stack PSFs with the EE from deep PSFs, which have higher signal-to-noise at large radii (but are saturated in the core) and with the optical model PSF. 

\section{Data}
\label{sec:Data}
In this study, we leverage two separate types of drizzled PSFs (``short-stack" and ``deep PSFs") to achieve high-fidelity measurements of the EE within different radii. The short-stack PSFs allow precise measurements of the EE at small radii, where the deep PSF images are saturated. The deep PSFs, on the other hand, provide much better signal-to-noise at large radii (beyond 2\unit{\arcsecond}), where the short-stack PSF signal is limited. For each of these, we compare the EE as a function of radius with the results from an optical model of the UVIS PSF \citep{2009wfc..rept...38H}. 

\vspace{0.5 cm}
\begin{figure}[ht]
    \centering
    \includegraphics[width=15cm]{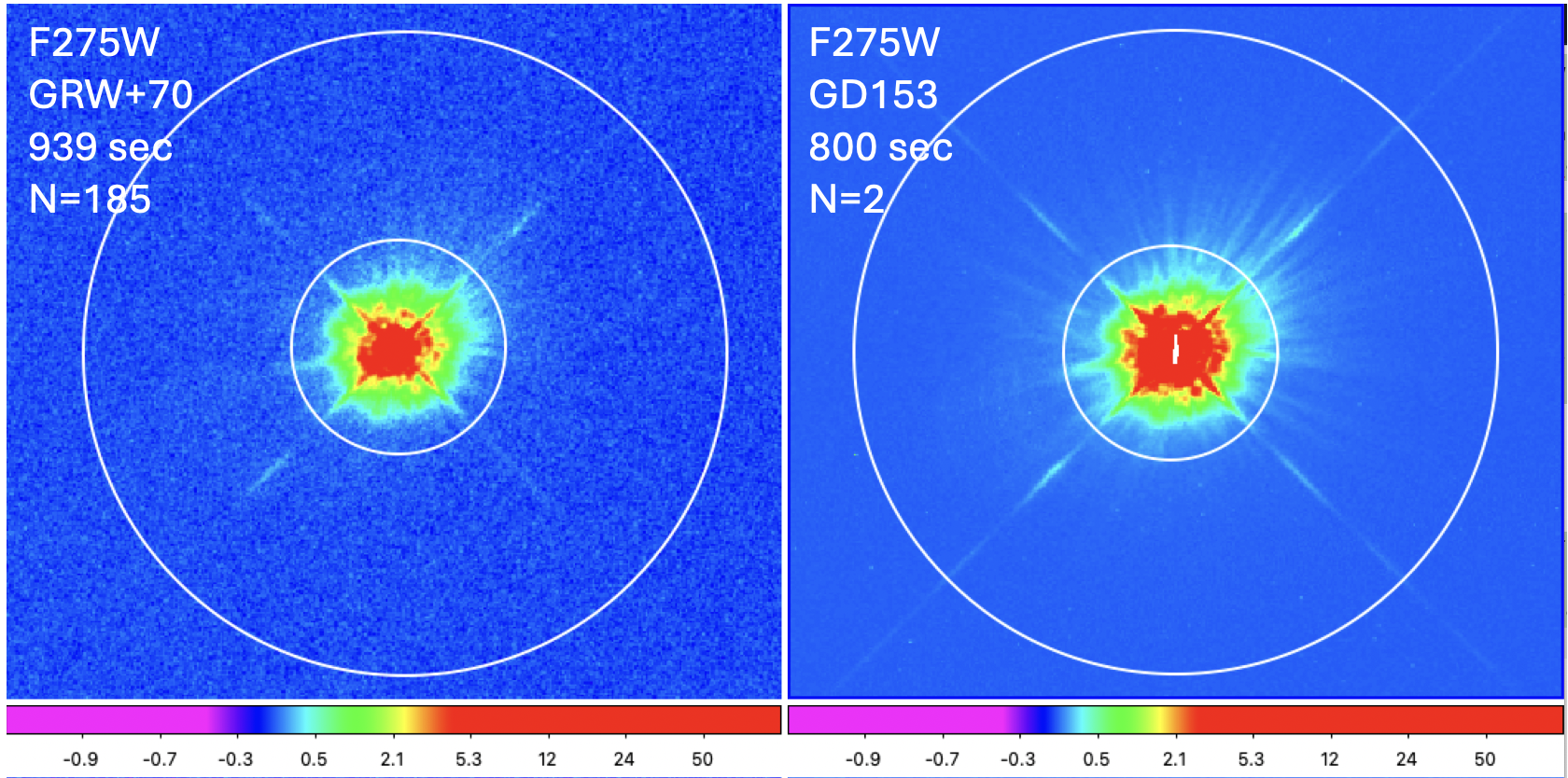}
    \caption{PSFs for F275W from short stack DRCs (left) and from deep PSFs (right). White circles indicate aperture radii of 2\unit{\arcsecond} and 6\unit{\arcsecond}.}
    \label{fig:PSFs_275}
\end{figure}
\vspace{0.5 cm}

\subsection{Short-Stack PSFs}
\label{sec:short-stack}

We recompute EE solutions using the drizzle-combined images of the CALSPEC standard stars GRW+70, GD153, and P330E used in the current UVIS photometric calibration by \citep{2022AJ....164...32C}. These ``short-stack" PSFs were created from calibrated FLC images obtained between 2009 and 2019. The science data arrays of individual exposures were rescaled to correct for changes in the observed count rates due to time-dependent sensitivity losses. Images were aligned in detector coordinates and combined with AstroDrizzle to produce high signal-to-noise PSFs out to a radius of 2\unit{\arcsecond}. Table 1 summarizes the short-stack observations used in this analysis.

\begin{minipage}[!ht]{1.0\linewidth} \vspace{3ex}
\normalsize
    \centering 
    \def\arraystretch{1.1} 
    \begin{tabular}{|l|c|c|c|r|r|c|} \hline
    
\textbf{Filter}  & \textbf{Amp}  & \specialcell{\textbf{Pivot}\\ \AA} & \specialcell{\textbf{Short Stack} \\ Target}  &  \textbf{N} & \specialcell{\textbf{Time} \\ (sec)} & \textbf{Epoch}   \\ \hline \hline
    F218W & A &  2228& GD153  &  8  & 57.6   &2016    \\ 
          & C &      &   &  6  & 74.4   &        \\ \hline 
    F225W & A &  2372& GD153  &  2 & 111.6   &2016    \\ 
          & C &      &   &  1  &  59.0   &   \\ \hline 
    F275W & A & 2710 & GRW+70 & 229 & 1012.2 &2020    \\  
          & C &      &  & 185 & 938.5  &     \\ \hline 
    F336W & A & 3354 & GRW+70 & 139 & 416.3 &2020  \\  
           & C &     &  & 115 & 374.1 &     \\ \hline \hline 
    F200LP & A & 4972 & P330E &  12  & 8.9  &2020   \\  
           & C &      &  &   6  & 5.0  &     \\ \hline 
    F350LP & A & 5874 & GD153 &  28 & 27.5  &2020  \\ 
           & C &      &  &   6 &  6.0  &       \\ \hline \hline 
    F475W* & A & 4773 & G191  &  22 & 13.1  &2016   \\ 
           & C &      &   &   7 &  5.1  &     \\ \hline
    F555W* & A & 5308 & G191  &  22 & 13.1  &2016     \\  
           & C &      &   &   7 &  5.1  &    \\ \hline  
    F606W* & A & 5889 & P330E &   14 & 23.0  &2016  \\ 
           & C &      &  &  12 & 21.0  &      \\ \hline 
    F625W* & A & 6243 & GD153 &   9 & 24     &2016   \\  
           & C &      &  &   6 & 30     &     \\ \hline \hline 
    F775W & A & 7651 & GD153  &   7 & 60.0  &2020   \\ 
          & C &      &  &   8 & 108.0 &       \\ \hline
    F814W & A & 8039 & GRW+70 & 117 & 1497.2 &2020  \\ 
          & C &      & & 134 & 736.0  &     \\ \hline
    F845M & A &  8439& P330E &  6 & 67.2   &2016       \\ 
          & C &      &  &  6 & 67.2   &        \\ \hline 
    F850LP & A & 9176 & GRW+70 & 36 & 1184.0 &2020 \\ 
          & C &      &  & 36 & 1184.0 &        \\ \hline 
    \end{tabular}
    
\captionsetup{type=table}
\captionof{table}{Short stack drizzled PSFs for a subset of UVIS filters. Columns provide the target, number of images, combined exposure time, and epoch. A few VIS filters were tested for validation (see asterix), but require no updates.}
    \label{tab:shortstack_nodeep}
\vspace{2ex}
\end{minipage}

\newpage

Note that 2016 short-stack PSFs are DRZ images (like DRC, but derived from FLT images with no pixel-based CTE-correction), but the CTE impact on the 2016 short-stack DRZ images in the detector subarray corners is negligible compared to the total flux of the CALSPEC standards. Given the absence of 2020 DRCs, their 2020 EE values were derived by scaling the 2016 results down by $\sim1\%$, the same factor as F275W (in the case of the UV filters), or by adopting the F814W EE correction (for all redder filters). This was based on an assumption of how applying time-dependent sensitivity corrections to the short-stack PSFs would affect the EE for similar wavelength filters. Given that we are redelivering corrected versions of the F275W and F814W EE curves, we revisit the 2016 DRZ images for UV and red filters, using the same methodology for filters with 2020 DRC images.

Figure \ref{fig:PSFs_275} compares the F275W short stack PSF of GRW+70 (N=185 frames, 939 seconds) and the deep PSF (N=2 frames, 800 seconds). For each PSF, Astrodrizzle was used to correct for distortion, match the sky background, flag bad pixels, and combine the individual input frames. 

\subsection{Deep PSFs}
\label{sec:deep-PSF}

At the beginning of WFC3's lifetime aboard the Hubble Space Telescope (HST), deep calibration images in two UVIS filters were acquired in program 11438 to characterize the fraction of light in the wings of the PSF at large radii. These were used to create a model of the UVIS EE across apertures and wavelengths \citep{2009wfc..rept...38H}. Bracketed short and long exposures in two filters, F275W and F625W, allowed for measurement of the PSF wings out to radius of $\sim 6$\unit{\arcsecond} (150 pixels) via the azimuthally-averaged fractional count rate per pixel and serve as a baseline for the optical model. These images provide excellent S/N at large radii, which we leverage in this analysis. 

\begin{minipage}[!ht]{1.0\linewidth} \vspace{3ex}
\normalsize
    \centering 
    \def\arraystretch{1.3} 
    \begin{tabular}{|l|c|c|c|c|r|r|} \hline
    
\textbf{Filter}  & \textbf{Amp}  & \textbf{Pivot} & \textbf{Target} &  \textbf{N} & \specialcell{\textbf{Time (sec)}} 
&\textbf{Proposal}  \\ \hline \hline
    F225W & C     &  2372& GRW+70 & 2 & 1120 & 17271 \\ \hline 
    F275W & ABCDM &  2710& GD153  & 2 &  800 & 11438 \\ \hline 
    F336W & C     &  3354& GRW+70 & 2 & 1000 & 17271 \\ \hline 
    F555W & C     &  5308& GRW+70 & 2 &  500 & 17271  \\ \hline 
    F625W & ABCDM &  6243& GD153  & 2 &  800 & 11438 \\ \hline 
    F814W & C     &  8039& GRW+70 & 4 & 1500 & 17271  \\ \hline 

    \end{tabular}
    
\captionsetup{type=table}
\captionof{table}{Observations used to create the ``deep PSFs" for 6 UVIS filters. Columns list the target, number of images, combined exposure time, and proposal ID used to estimates the EE at large radii.  Program 11438 observed two filters at 5 positions each, near the center of each amp (A, B, C, D) and near the middle (M) of the FoV. Program 17271 placed the target in same amp C position as in program 11438 to observe PSFs in additional filters.}
    \label{tab:Deep_PSF_info}
\vspace{2ex}
\end{minipage}

\vspace{0.5 cm}
The deep PSF images in program 11438 were acquired at 5 positions: near the center of each amplifier (A, B, C, and D) and near the middle of the FoV (M). This allows us to look for any variations in the EE at large radii due to changes in the angle and position of the UVIS filter ghosts. Additional deep PSF images were acquired recently in program 17271 (PI: Bajaj) to sample additional UVIS filters (F225W, F336W, F555W, F814W) across the entire range of UVIS wavelengths. The observations use the same commanded aperture (UVIS-CENTER) and large dither (-51\unit{\arcsecond}, -55\unit{\arcsecond}) to place the star in the same amp C position observed in program 11438. PSFs in the F438W filter suffered a guide star failure and were not included in the analysis. Table \ref{tab:Deep_PSF_info} summarizes the two calibration programs used to produce deep PSFs, including the number of input exposures for each filter and the combined exposure time.

\vspace{1 cm}
\begin{figure}[ht]
    \centering
    \includegraphics[width=16cm]{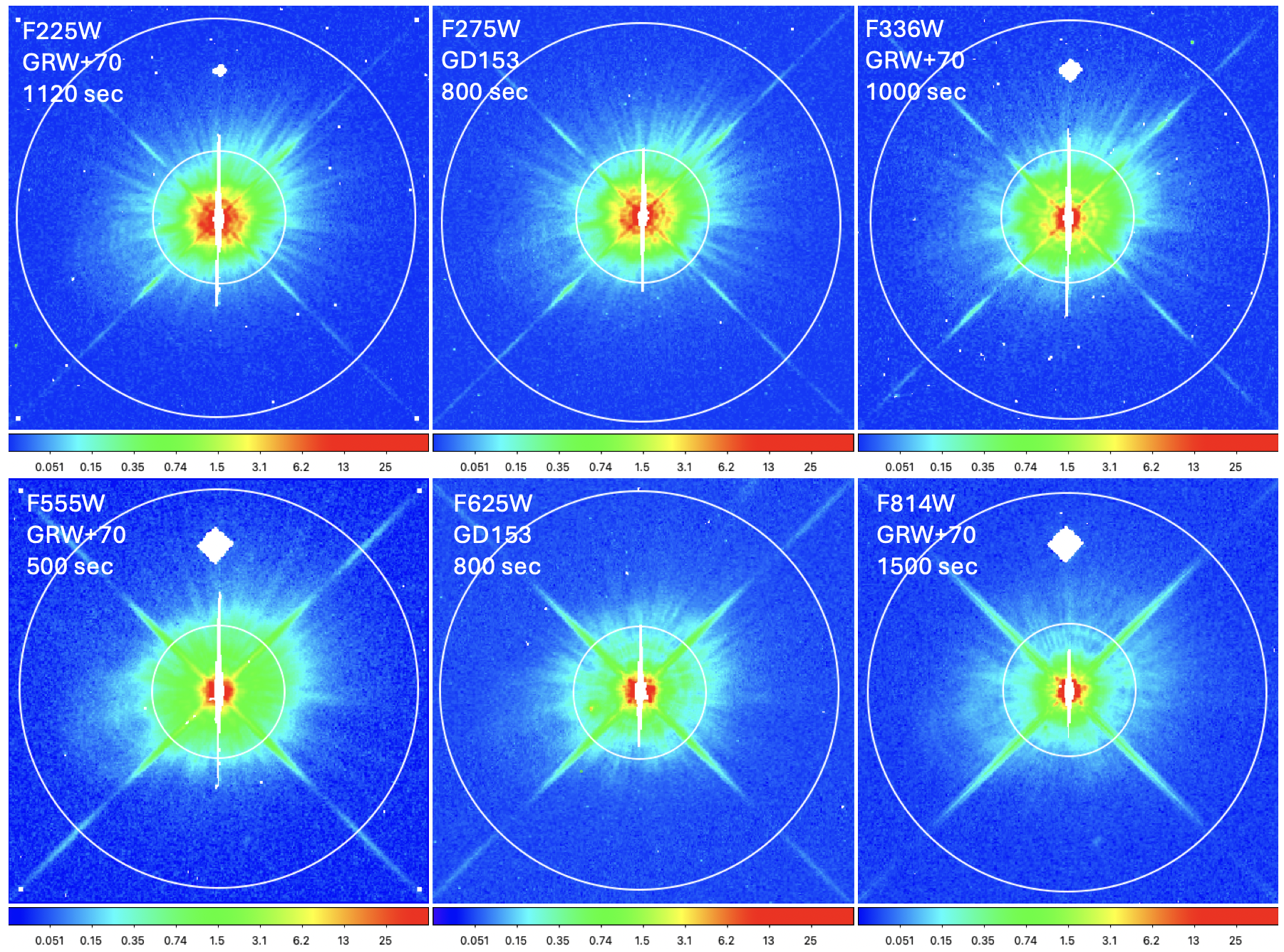}
    \caption{Deep PSFs in six UVIS filters in the center of amp C.  White circles indicate aperture radii of 2\unit{\arcsecond} and 6\unit{\arcsecond}. For the F275W and F625W filters, deep PSFs are based on GD153. For the other four filters, deep PSFs are based on GRW+70 and include a mask for the red companion star at a radius of 4.5 arcsec. Saturation bleeds for each star are also masked. See Section \ref{sec:deep-PSF} for details.}
    \label{fig:Deep_PSFs}
\end{figure}

\vspace{0.5 cm}

To construct the deep PSFs, we retrieved calibrated FLT/FLC data products for these two programs from MAST and corrected for cross-talk using the Python algorithm provided on the WFC3 website \footnote{https://www.stsci.edu/hst/instrumentation/wfc3/software-tools/crosstalk}. (Only PSFs at the middle of the FoV suffered from cross-talk within the 6\unit{\arcsecond} aperture). AstroDrizzle was used to flag cosmic rays and then `grow' their data quality flags, including saturation bleeds, before combining the individual exposures. Pixels with no valid input data were assigned NaN values in the drizzle-combined image. For GRW+70, the faint red companion at $\sim$4.5\unit{\arcsecond} separation was also masked in the individual input frames before drizzling. Finally, any NaN pixels in the deep PSFs were replaced by the local median in a 3×3 box. Figure \ref{fig:Deep_PSFs} shows the deep PSF observations, prior to NaN-filling. 
\section{Constructing the EE Curves}
\label{sec:construct_EE_curve}

\subsection{Comparing the Drizzled PSFs }

Reliable empirical EE measurements require that the signal in the PSF remains measurable above the noise out to a radius of 6\unit{\arcsecond}. However, a closer examination of the short-stack images reveals that the normalized annular azimuthal average\footnote{The count rate is normalized by the total count rate within 6\unit{\arcsecond} aperture as given by \href{https://www.stsci.edu/hst/instrumentation/reference-data-for-calibration-and-tools/synphot-throughput-tables}{\texttt{synphot}}.} becomes noisy beyond $\sim 50$ pixels (2.0\unit{\arcsecond}) for most filters, even in cases where additional exposures were added in 2020. While a few filters (e.g. F275W, F814W, and F850LP) retain measurable signal out beyond 2\unit{\arcsecond}, the rapid decline in signal-to-noise at larger radii precludes reliable empirical EE measurements to 6\unit{\arcsecond} for any UVIS filter.

Azimuthally averaged profiles from short-stack PSFs are provided in Appendix D (Figures \ref{fig:LP_SN}–\ref{fig:VIS_SN}) for LP, UV, and VIS filters. Figure \ref{fig:az_avg_key_filts} shows the wavelength dependence of the UVIS PSF as a function of aperture radius, including a prominent halo in F275W between 10 and 30 pixels that becomes weaker and shifts to larger radii in redder filters.

\begin{figure}[!ht]
    \centering
    \includegraphics[width=17cm]{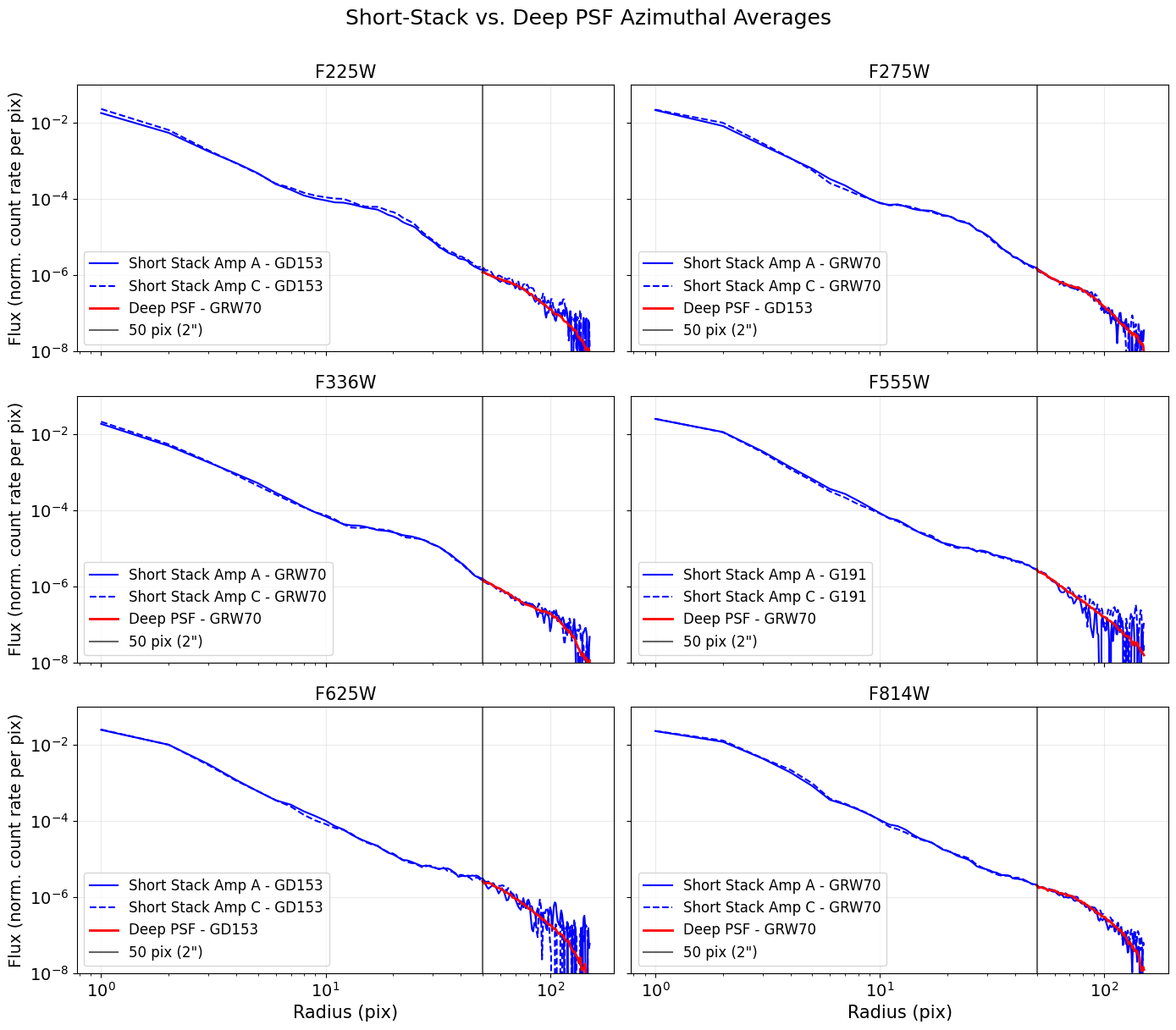}
    \caption{Azimuthal average flux (in units of count rate per pixel) vs radius for PSF observations in the center of amp C in six filters (F225W, F275W, F336W, F555W, F625W, and F814W). Short stack PSFs from 2016 and 2020 are overplotted with the deep PSFs, and show very close agreement between 2" (vertical line) and 6".}
    \label{fig:short_vs_deep}
\end{figure}

Six filters have deep PSF data, which we use to constrain the fraction of light between 2\unit{\arcsecond} and 6\unit{\arcsecond}. Figure \ref{fig:short_vs_deep} compares the azimuthally averaged count-rate profiles of the short-stack and deep PSFs. The 2020 short stacks (F275W, F336W, and F814W), constructed from hundreds of input exposures, retain higher signal-to-noise at large radii than the 2016 short stacks (e.g., F225W, F555W, and F625W), but in all cases the deep PSFs provide superior signal-to-noise between 2\unit{\arcsecond} and 6\unit{\arcsecond}. Azimuthal averages are not computed for the deep PSFs at radii smaller than 2\unit{\arcsecond}, where saturation spikes compromise measurements of the mean annular count rate.

For GRW+70 a faint, red companion star at a radius of 4.5\unit{\arcsecond} (110 pixels) is masked in the deep PSF frames before stacking, and no trace is visible in the azimuthally averaged profiles. No masking was applied to the faint companion star in the short stack images; however, since the images were aligned in detector coordinates (and not in sky coordinates), the companion rotates around the target star and its signal is too weak to be detectable above the noise in the azimuthally averaged profiles.

For filters with deep PSFs, the high-S/N empirical measurements of the PSF wings can be used to directly scale and replace the short-stack EE curves outside of 2\unit{\arcsecond}. We therefore apply a scalar normalization to each 1D short-stack EE curve so that its EE value at 2\unit{\arcsecond} matches the empirical EE measured from the deep PSF. To compare the short-stack and deep PSFs, we first normalize the count rates of each star at different aperture radii by the total synthetic count rate of the star at a radius of 6\unit{\arcsecond}, as predicted by \href{https://www.stsci.edu/hst/instrumentation/reference-data-for-calibration-and-tools/synphot-throughput-tables}{\texttt{synphot}}, using the UVIS throughput curves and the latest spectrum in the CALSPEC database for each star. The deep PSF and short-stack azimuthal average profiles show good agreement at 2\unit{\arcsecond} radius; however, since deep PSFs are only available for 6 filters, we assess the performance of the optical model at 2\unit{\arcsecond} against the deep PSFs with the goal of using the optical model to scale EE curves in the remaining filters. 

The narrow-band filters F280N and F953N (both with 2016 short-stack PSFs) become very noisy well before $r = 50$ pixels (see Figures \ref{fig:UV_SN}, \ref{fig:VIS_SN}). The methodology described in this report is therefore insufficient for these datasets, and we leave their EE values unchanged. In future work, we plan to create deeper short-stack PSFs for more filters (including F280N and F953N) in order to achieve adequate signal-to-noise with measurability to 2\unit{\arcsecond} and beyond.

\subsection{Leveraging the Optical Model}
\label{sec:lev_optical_model}
From 2009 to 2016, an optical model of the UVIS PSF by \citet{2009wfc..rept...38H} was used to estimate the EE versus radius for all 42 UVIS imaging filters. 
In 2016, the models were replaced with empirical EE measurements derived from short-stack drizzled PSFs. The empirical data were normalized to the model EE values at a radius of 1.4\unit{\arcsecond}, where the signal-to-noise of the empirical data drops off rapidly (e.g. see Figure 7 of \citep{2016wfc..rept....3D}.

We similarly explore using the optical model to normalize the short-stack EE curves where the empirical measurements become noise-dominated. However, the optical model provides tabulated values at discrete radii, e.g. 1.0\unit{\arcsecond}, 1.5\unit{\arcsecond}, and 2.0\unit{\arcsecond}, but not 1.4\unit{\arcsecond}. The model EE at 1.4\unit{\arcsecond} must therefore be obtained by linear interpolation between the tabulated values at 1.0\unit{\arcsecond} and 1.5\unit{\arcsecond}. Since the EE is not a linear function of radius, this introduces a small systematic offset relative to the true model value at 1.4\unit{\arcsecond}. Normalizing the short-stack EE curves at 2\unit{\arcsecond} instead anchors them to a tabulated model value and therefore avoids this small systematic error.

\cite{2009wfc..rept...38H} notes that the optical model over-predicts the EE for F275W at 2\unit{\arcsecond} (a similar problem as for ACS/HRC.)  We thus explore the accuracy of the optical model by revisiting deep PSF data used as the basis of the model, along with deep PSF data in a few other key filters, to check the fraction of light between 2\unit{\arcsecond} and 6\unit{\arcsecond}. 

\begin{figure}[ht]
    \centering
    \includegraphics[width=15cm]{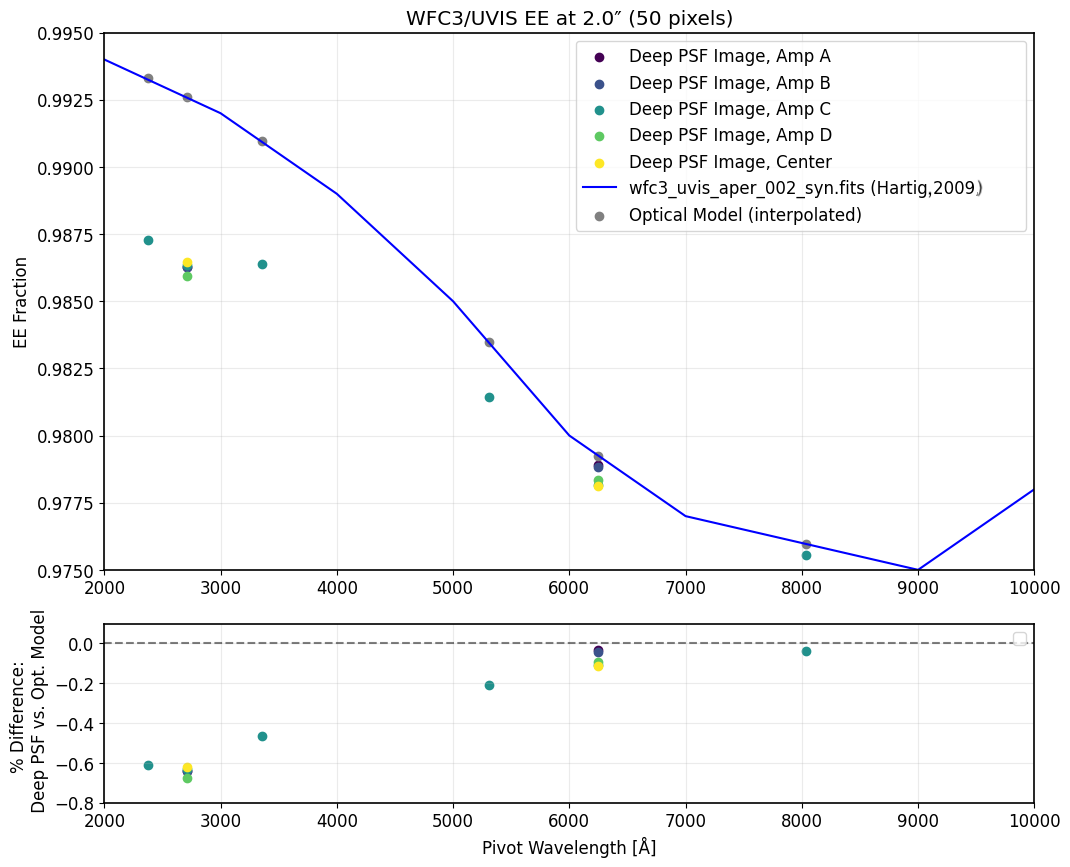}
    \caption{Encircled energy fraction at 2.0" (50 pixels) as a function of filter pivot wavelength. The solid line shows EE values from the 2009 optical model, interpolated by filter pivot wavelength, while colored points show the empirical results from deep PSFs in six filters. F275W and F625W were observed at 5 positions across the FoV and show a small scatter of $\sim$0.1\% due to changes in position of the filter ghosts and uncertainties in the flat field.}
    \label{fig:model_vs_deep}
\end{figure}

Figure \ref{fig:model_vs_deep} compares the EE at 2\unit{\arcsecond} vs. wavelength from the optical model and from deep PSFs in 6 filters. F275W and F625W were measured at 5 locations across the detector and show a small intrinsic scatter of $\sim$0.1\%, where the others were only from amp C. We find excellent agreement between the optical model and the empirical PSFs for VIS filters (within 0.1 - 0.2\%). For UV filters F225W, F275W, and F336W, however, the model over-estimates the observed fraction of light at 2\unit{\arcsecond} by $\sim 0.5 \pm  0.1$\%. 

\subsection{Merging the EE Results}

Given the lack of signal at large radii in the short-stack PSFs, the measured EE curves cannot be reliably normalized to their empirical count rates at 6\unit{\arcsecond}. Instead, we adopt a reference EE value at 2\unit{\arcsecond}, derived either from the optical model or from the deep PSFs. For each filter, we apply a scalar normalization to the entire short-stack EE curve such that its EE value at 2\unit{\arcsecond} matches the adopted reference value. The shape of the empirical EE curve interior to 2\unit{\arcsecond} is therefore preserved, while the absolute normalization is anchored by the reference EE at 2\unit{\arcsecond}. 

For UV filters (F225W, F275W, and F336W), the deep PSFs provide a better estimate of the EE beyond 2\unit{\arcsecond} compared to the optical model, which over-estimates the encircled energy. We therefore normalize the short-stack EE curves to match the deep PSF EE value at 2\unit{\arcsecond}. Because the deep PSFs are saturated at small radii, these are used only to set the normalization and to constrain the EE between 2\unit{\arcsecond} and 6\unit{\arcsecond}; the shape of the EE curve at smaller radii is determined from the short-stack PSFs. EE curves from deep PSFs are normalized using the synthetic count rate of the star\footnote{The synthetic count rates were computed using \href{https://www.stsci.edu/hst/instrumentation/reference-data-for-calibration-and-tools/synphot-throughput-tables}{\texttt{synphot}}.} before computing the EE.

For VIS filters with deep PSFs (F555W, F625W, and F814W), the model EE at 2\unit{\arcsecond} is consistent with the empirical EE to within the measurement uncertainty of $\sim0.1\%$ (see Figure \ref{fig:model_vs_deep}). We therefore use the optical model EE at 2\unit{\arcsecond} as the normalization reference for VIS filters, scaling the entire short-stack EE curve to match the model at this radius.
For F555W, we note that the optical model differs from the deep PSF empirical measurement by about $0.2\%$ at 2\unit{\arcsecond} (compared to a $0.1\%$ measurement uncertainty). Though we scale the F555W EE (along with nearby filters) to the optical model for this report, we note that the F555W, F475W, F606W, and F625W EE curves are NOT UPDATED in this delivery. These four filters were included only to validate the updated methodology across the entire UVIS wavelength range. 

\section{Results}
\label{sec:results}

\subsection{Final EE Curves}

\begin{figure}[h]
    \centering
    \includegraphics[width=16cm]{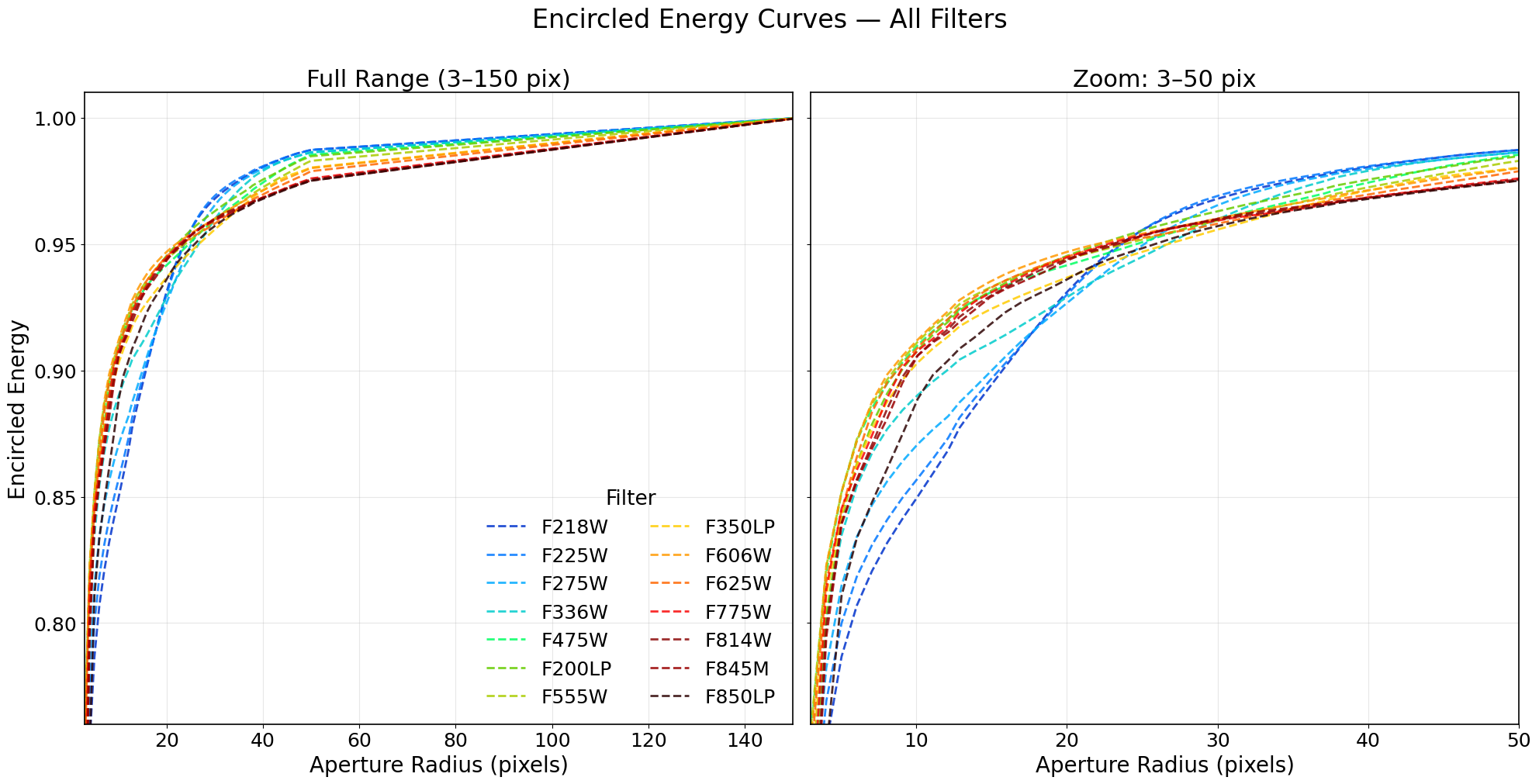}
    \caption{New encircled energy curves, color coded by pivot wavelength (blue to red), for a subset of UVIS filters discussed in this report.  (Left) The EE versus radius from 3 to 150 pixels (0.12" to 6.0") and (right) a zoomed version from 3 to 50 pixels (0.12" to 2.0").
}
    \label{fig:EE_curves}
\end{figure}

New EE curves from 3 to 150 pixels are presented in Figure \ref{fig:EE_curves} for Amp C (UVIS2), with a zoomed version from 3 to 50 pixels (0.12\unit{\arcsecond} - 2.0\unit{\arcsecond}). The 14 filters discussed in this report (10 of which will be included in the delivery) are color-coded from blue to red by pivot wavelength. A clear wavelength-dependent trend in the EE curve shapes is apparent. 

\vspace{0.3 cm}

\vspace{0.3 cm}

The updated EE curves show the same wavelength-dependent behavior reported by \citet{2022wfc..rept....2M}: blue filters have lower EE at small radii, while redder filters contain a larger fraction of their total flux in the PSF core. At larger radii, this trend reverses, with the EE in blue filters increasing more than in red filters. The long-pass filter F850LP remains a notable exception, with EE values at small radii that fall between the blue and red broadband filters before converging toward the F775W and F814W curves at larger radii.

Overall, these trends are consistent with Figure 6.15 in the WFC3 Instrument Handbook \citep{2025wfci.book...18M}, which shows EE curves at 200, 400, and 800 nm crossing near 0.9\unit{\arcsecond} (23 pixels) at an EE value of 0.95. However, this figure was created using the optical model, which over-estimates the UV EE at large radii. \cite{2009wfc..rept...38H} suggests this is due to ``spatial frequency limitations of the OTA mirror maps at UV wavelengths". An attempt is made to improve the optical model for UV filters by ``filtering the primary mirror OPD map in the frequency domain, emphasizing the power at higher frequencies", however ``this technique cannot add sufficient power at high enough frequencies to improve the fits at radii larger than $\sim1.2$\unit{\arcsecond}" \citep{2009wfc..rept...38H}. 

In Figure \ref{fig:EE_curves} of this report, the F275W and F814W EE curves cross at a slightly larger radius of $\sim 1.1$\unit{\arcsecond} (28 pixels) than predicted by the optical model, and this is consistent with Figure 9 of \cite{2022wfc..rept....2M}. We also note that the updated F200LP and F350LP EE curves now follow the expected pivot wavelength ordering more closely than in the 2020 EE curves.

\subsection{Improvements at small radii}

A key motivation of this study is to improve EE measurements for small aperture radii ($r \le 10$ pixels) and to compare our results with those of \cite{2025wfc..rept....5H} for aperture corrections between 5 and 10 pixels. Because the normalization of the EE curves (based on model or deep PSF EE at 2 \unit{\arcsecond}) is applied as a single multiplicative scaling, the relative shape of the short-stack EE curve interior to 2\unit{\arcsecond} is unchanged. As a result, aperture corrections between small radii (e.g., 5–10 pixels) depend only on changes to the measured curve shape, not on the choice of normalization. However, the adopted EE at large radii determines the total flux and therefore affects the photometric zeropoint.

 Systematic offsets of $\sim1$ pixel are found between the PSF centroids derived here and those from the prior analysis for some filters. At small radii, these offsets have a substantial effect on the EE. As described in Section \ref{sec:methodology}, refined centroids are obtained for each PSF in the short-stack images by fitting a 2D Gaussian profile within a small cutout centered on the initial centroid, allowing us to find a new centroid with sub-pixel accuracy. We assess the combined impact of improved centroiding and the revised normalization at 2\unit{\arcsecond} on the delivered EE curves and zeropoints. For each filter, we separately quantify changes to the EE at r=5 and r=10 pixels. These are provided in Appendix C in Tables \ref{tab:r5_changes} and \ref{tab:r10_changes}. Figures \ref{fig:EE_at_5} and \ref{fig:EE_at_10} plot the EE at r=5 and r=10 pixels after implementing improved centroids and normalization (pinning to either the optical model or to deep PSF data outside of 2\unit{\arcsecond}).  At r = 5  pixels, the updated EE values differ from the 2020 values by up to 0.034 mag for wide filters and 0.044 mag for LP filters, while at r = 10 pixels the new values differ from the 2020 values by up to 0.012 mag for wide filters and 0.019 mag for LP filters, though most re-delivered filters change by 0.005 mag or more for at least one chip (see Table \ref{tab:r10_changes}).

In three filters (F275W, F336W, and F814W) the updated EE curves yield aperture corrections that are in substantially better agreement with the focus-diverse PSF-based measurements of \cite{2025wfc..rept....5H}.

\begin{figure}[!t]
    \centering
    \includegraphics[width=16.5cm]{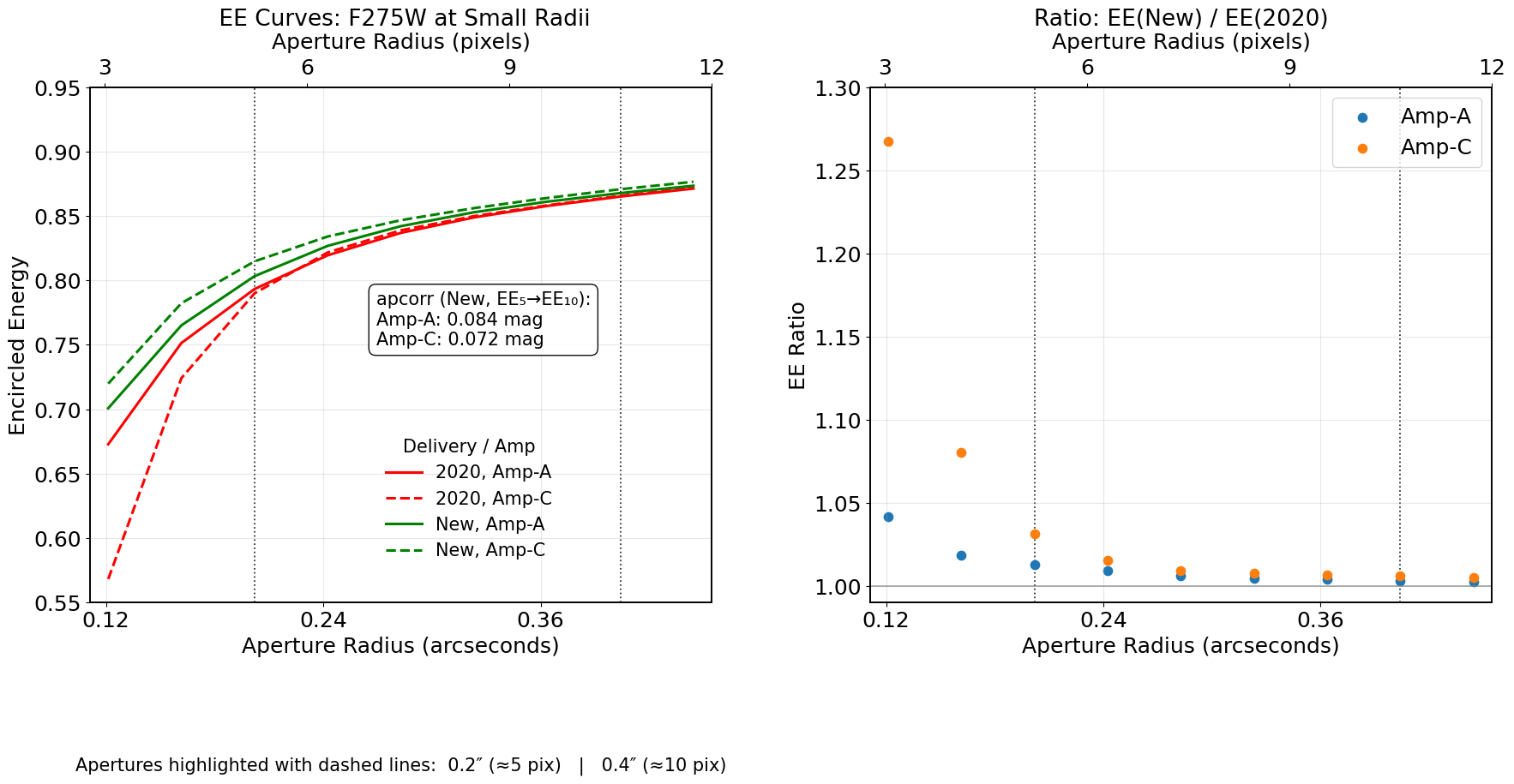}
    \caption{(Left) F275W encircled energy for small aperture radii (less than 12 pixels) from the 2020 solutions (red) and the new solutions with improved centroids (green). Vertical lines indicate 5 and 10-pixel (0.2 and 0.4 arcsec) aperture radii. (Right) Ratio of new and 2020 EE values. Centroid improvements have a large impact on the EE at 5 pixels, for example, but only a small impact at 10 pixels where the zeropoints are computed. }
    \label{fig:F275W_centroid}
\end{figure}

\begin{figure}[!t]
    \centering
    \includegraphics[width=16.5cm]{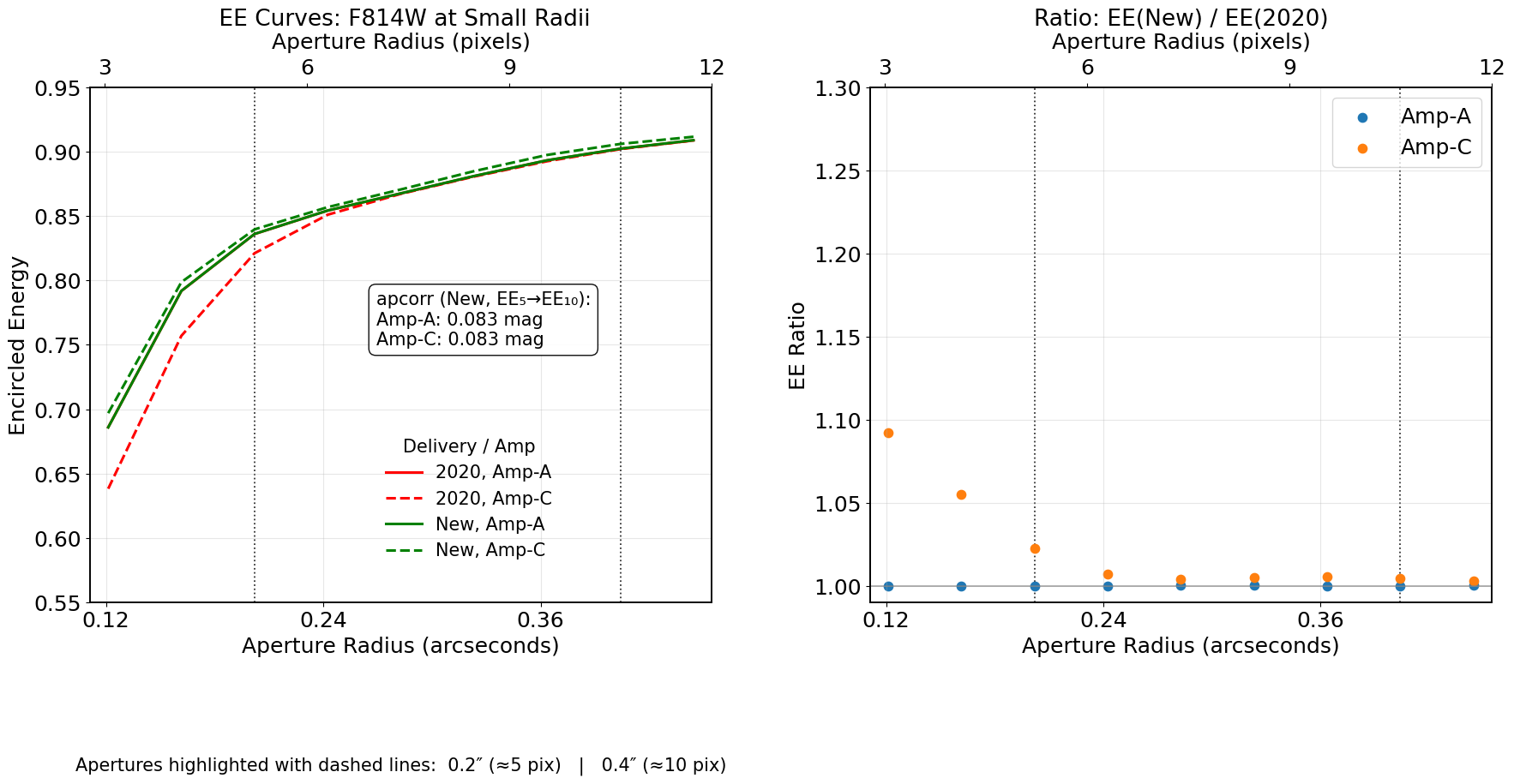}
    \caption{Same as Figure \ref{fig:F275W_centroid} except for the F814W filter.}
    \label{fig:F814W_centroid}
\end{figure}

\subsection{ Aperture Corrections between 5 and 10 pixels}
\label{sec:apcorr}

The primary difference in shape between the new EE curves and those from 2020 at small radii is the improved centroiding introduced in our updated methodology. By fitting a 2-D Gaussian to the star's PSF in the drizzled short-stack images, we obtain a more accurate center position, which improves the fidelity of EE measurements for radii less than 10 pixels and translates into revised aperture corrections between 5 and 10 pixels. Figure \ref{fig:F275W_centroid} and \ref{fig:F814W_centroid} show how the updated (2026) EE curves in F275W and F814W compare to the 2020 curves.

\begin{figure}[!b]
    \centering
    \includegraphics[width=15cm]{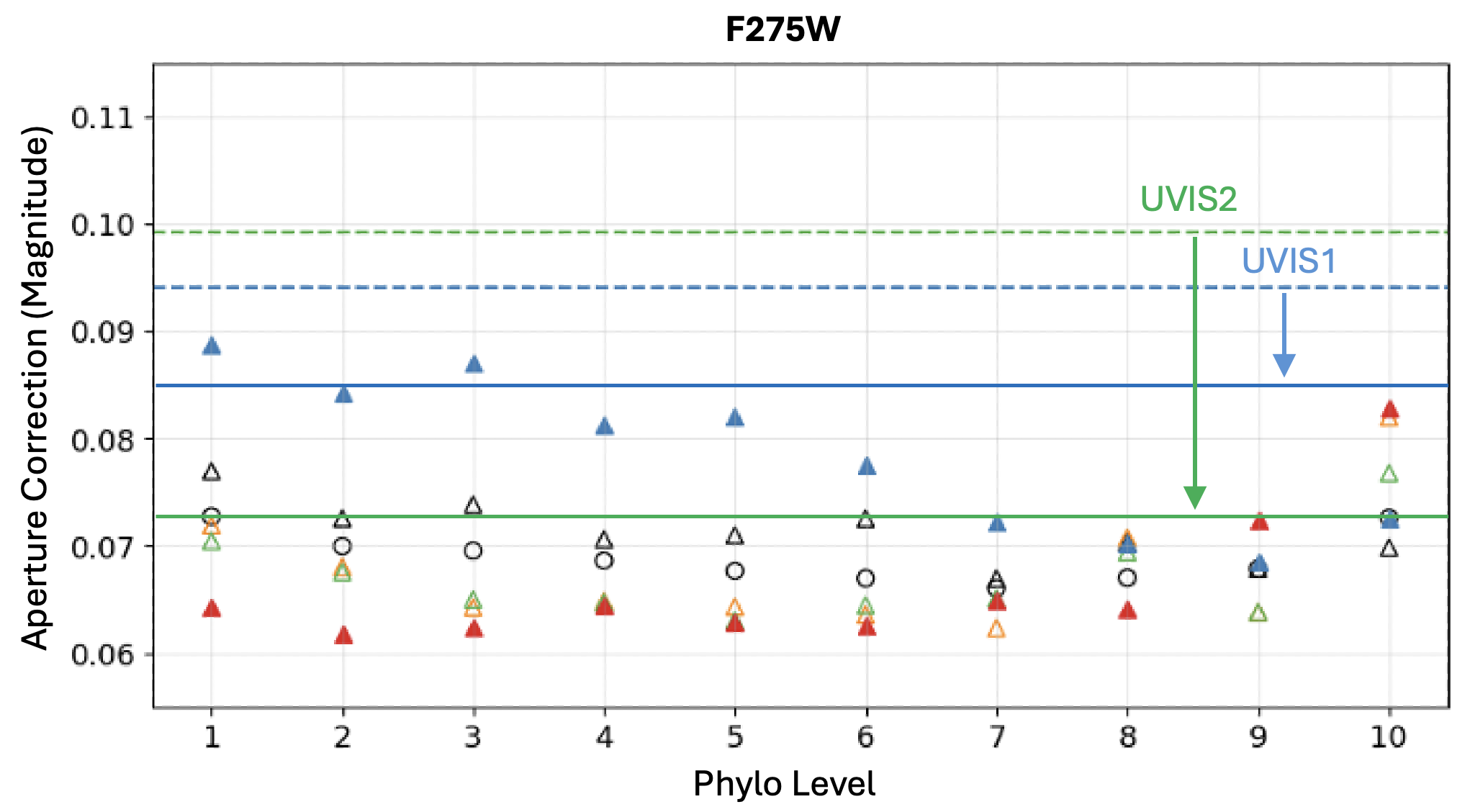}
    \caption{F275W aperture correction (mag, r = 5 - 10 pixels) vs. focus (phylo) level from archival PSFs (adapted from \cite{2025wfc..rept....5H}). 
    Blue and green triangles were derived from PSFs in the amp A and C corner subarrays; open triangles show the center of the detector; open circles show the mean over the FoV. 
    Dashed lines were computed using 2020 EE tables for UVIS1 amp A (blue) and UVIS2 amp C (green). 
    \textbf{Solid lines are based on the updated EE and are now consistent with the mean for each amp}. The UVIS2 value is similar to the mean over the FoV, while UVIS1 is $\sim0.01$ mag larger. 
}
    \label{fig:phylo_apcorr}
\end{figure}
\vspace{0.5 cm}

Importantly, the revised aperture corrections are in significantly better agreement (see Figure \ref{fig:phylo_apcorr}) with the focus-diverse PSF-based measurements reported in ISR 2025-05 \citep{2025wfc..rept....5H}, indicating that centroiding errors were a dominant term in the previously observed discrepancies. Table \ref{tab:compare_apcorr} compares the focus-diverse PSF-based mean aperture correction between 5 and 10 pixels from \cite{2025wfc..rept....5H} with those from the newly calculated EE values. Although short-stack PSFs for F336W were computed in 2020, the corresponding EE curves were never delivered, and so this filter has retained the 2016 EE values until now. Since our new analysis shows improved consistency with archival PSFs for UVIS1 at small aperture radii, we include F336W in the current update. For the two VIS filters F438W and F606W, no significant errors were found in the EE at small radii in \cite{2025wfc..rept....5H}, suggesting that the 2016 short-stack EE solutions were not affected by centroiding errors. Improving the centroiding in the 2020 short-stack EE measurements allows us to correctly capture more of the flux from the PSF at small radii.

Even using the new EE tables, aperture corrections derived from the EE tables are slightly larger than the values from archival PSFs. \cite{2025wfc..rept....5H} attribute this difference to the different sky annulus used for each method. While the 2020 EE tables measure the sky in an annulus between 160 and 200 pixels, the aperture correction maps were derived from shorter exposures and adopt a sky annulus between 15 and 20 pixels. This results in a small fraction of light in the PSF wings being subtracted.  As the 10 pixel aperture contains four times the number of pixels as the 5 pixel aperture, this systematic overestimation of the background causes underestimates of the flux in the larger aperture more so than in the 5 pixel aperture. This produces aperture corrections which are slightly smaller than predicted from EE tables. To estimate the effect of the different sky annulus, \cite{2025wfc..rept....5H} recalculate the F814W UVIS1 EE using a sky annulus from 15 to 20 pixels and show that this lowers the predicted correction by about 0.01 mag.

\vspace{0.3 cm}
\begin{minipage}[!hb]{1.0\linewidth} \vspace{3ex}
\normalsize
    \centering 
    \def\arraystretch{1.1} 
    \begin{tabular}{|l|c|c|c|c||r|} \hline
    
\textbf{Filter}  & \textbf{Amp}  & \textbf{Archival PSFs}
& \textbf{2026 Apcor} & \textbf{2020 Apcor} & \textbf{$\Delta Apcor$}   \\ \hline \hline

F275W  & A & 0.068 - 0.089 & 0.084 & 0.094 &         -0.010  \\  
       & C & 0.064 - 0.077 & 0.072 & 0.099 & \textbf{-0.027} \\ \hline \hline
F336W  & A & 0.080 - 0.096 & 0.083 & 0.093 & \textbf -0.010 \\ 
       & C & 0.069 - 0.080 & 0.070 & 0.068 &         +0.002  \\ \hline \hline
F438W* & A & 0.079 - 0.094 &  -    & 0.084 &           -    \\ 
       & C & 0.065 - 0.074 &  -    & 0.070 &           -    \\ \hline \hline
F606W* & A & 0.077 - 0.089 &  -    & 0.085 &           -    \\ 
       & C & 0.072 - 0.083 &  -    & 0.076 &           -    \\ \hline \hline
F814W  & A & 0.070 - 0.088 & 0.083 & 0.082 &         +0.001  \\ 
       & C & 0.073 - 0.086 & 0.083 & 0.102 & \textbf{-0.019} \\ \hline 

    \end{tabular}
    
\captionsetup{type=table}
\captionof{table}{Aperture correction from r = 5 to 10 pixels (in magnitudes) for five UVIS filters with focus-diverse PSFs from Anderson (2018). For each CCD amplifier, column 3 gives the range of aperture corrections derived from a large sample of archival PSFs (Huynh et al. 2025), column 4 gives the value derived from the 2026 EE tables, while column 5 was derived from the 2020 EE tables. The last column gives the difference $\Delta Apcor$ (2026 - 2020), with changes exceeding 0.010 mag (in absolute value) in bold. (Filters with an asterix are derived from 2016 short stacks and are consistent with the archival PSFs between 5 and 10 pixels.)}
    \label{tab:compare_apcorr}
\vspace{2ex}
\end{minipage}

Noting the dependence of the EE results on the sky annulus, we compare the method used to compute the background for 2020 EE tables and for the 2020 photometric zeropoints. For the EE, short stack PSFs used the sigma-clipped mean in an annulus with radii between \textbf{160 and 200 pixels} (\cite{2022wfc..rept....2M}), whereas the zeropoints used the sigma-clipped mean in an annulus between \textbf{156 and 165 pixels}. For consistency with the 2020 zeropoints, we adopt the same sky annulus (from 156 to 165 pixels) for the new EE analysis in order to reduce any (small) systematic errors in the photometric calibration.

\section{Discussion}
\subsection{Addressing Discrepancies with 2020 EE Curves}

At small radii, such as 0.2\unit{\arcsecond} and 0.4\unit{\arcsecond} (5 and 10 pixels), the updated centroids in the short-stack PSFs correct for systematic errors in the PSF centroids of up to 1 pixel for some filters. These lead to significant improvements in encircled energy for small apertures, where the EE is most sensitive to centering errors. 

Additionally, the absolute value of the EE is affected by the adopted normalization. For most filters in this delivery, the EE at 0.4\unit{\arcsecond} changes by at least 0.005 mag for one chip, typically UVIS2. The largest differences are found in F218W (-0.012 mag for UVIS2) and F850LP (+0.019 and +0.016 mag for UVIS1 and UVIS2, respectively). These changes do not follow a simple pattern because the 2020 EE delivery relied on several filter-specific assumptions. The discussion below summarizes those assumptions and their impact on the revised EE values.

In 2020, the F275W EE curves showed closer agreement at 10 pixels for UVIS1 and UVIS2 and better agreement with the optical model compared to the 2016 EE. Following these results, EE values for the other UV filters with 2016 PSFs (F218W, F225W, F280N) were also changed by $\sim1\%$ at 10 pixels by multiplying by a scalar correction. For F218W and F225W, the UVIS1 values were adopted for UVIS2. Updated EE curves were recomputed for F336W, but were not delivered and therefore retained the 2016 EE solution in the 2020 delivery. 

For F814W, the 2020 analysis of the PSFs produced EE values which were closer to the model and nearly identical between 7 and 12 pixels for the two chips. For longer wavelength filters, the empirical PSFs showed large deviations from the optical model EE at 10 pixels. Because these filters were based on fewer datasets, the F814W values for each CCD were adopted for F845M and F850LP. For F775W, the EE values at 10 pixels for UVIS2 were $\sim$0.5\% larger than for UVIS1, so the UVIS1 values were adopted for UVIS2.

In our new analysis, we apply a consistent methodology across filters and chips, in which we empirically measure the EE out to 2\unit{\arcsecond} and subsequently scale the EE curves to either the optical model or the deep PSF EE value at 2\unit{\arcsecond}.  

\subsection{EE Differences Between UVIS1 and UVIS2}
\label{sec:EE_diff}

\begin{figure}[h]
    \centering
    \includegraphics[width=16cm]{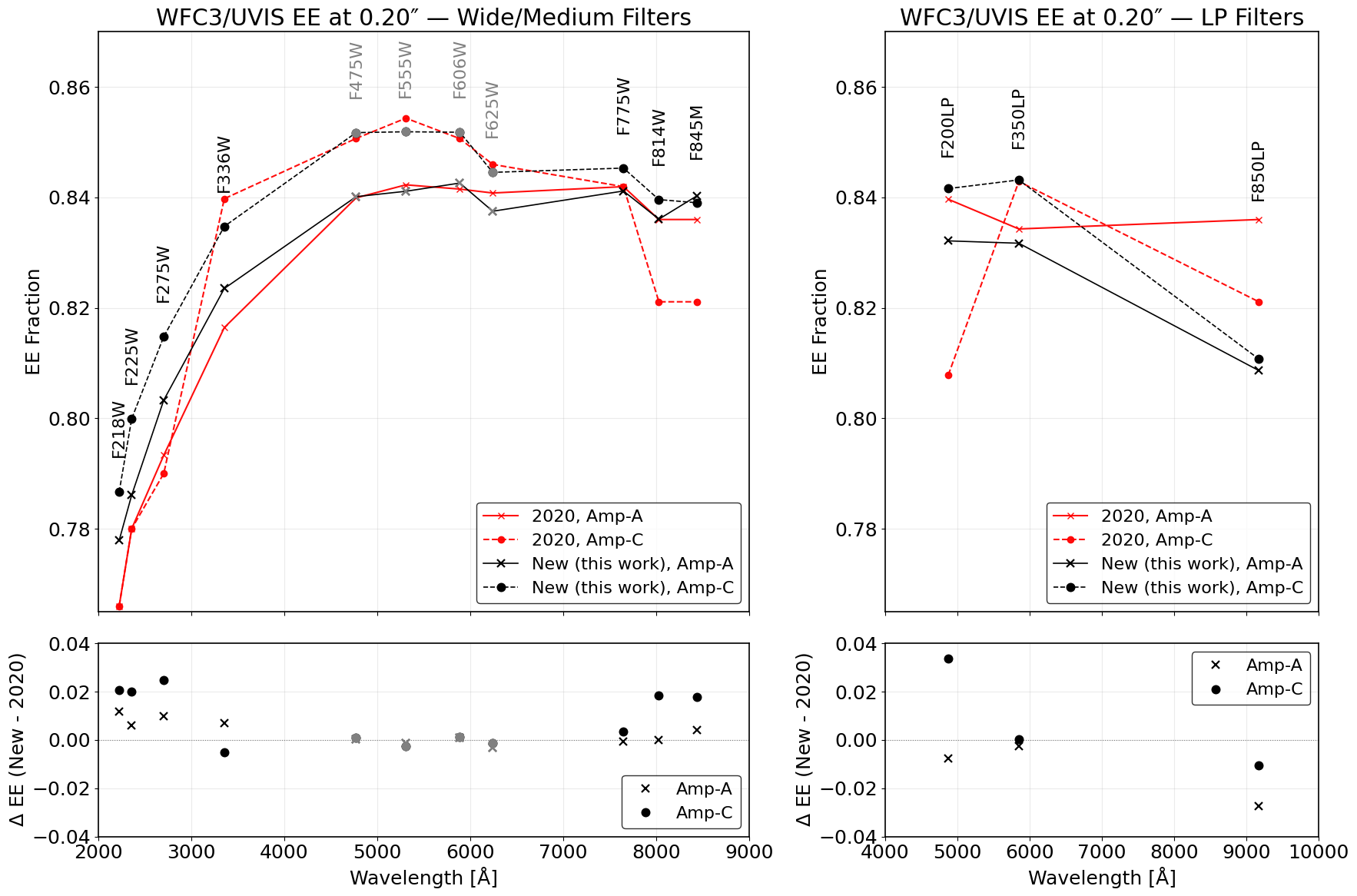}
\caption{(Top panels) Encircled Energy in a 5-pixel aperture radius, EE(5), versus pivot wavelength for wide and medium filters (left) and LP filters (right). Filters tested in this study are shown as X's for UVIS1 and circles for UVIS2. Black symbols indicate filters identified for updates, while gray symbols show additional filters examined. At 5 pixels, UVIS2 (dashed lines) has a noticeably larger EE fraction than UVIS1 (solid lines). (Bottom panels) Difference between the new EE(5) and the 2020 EE(5). Values are tabulated in Appendix C.
}
    \label{fig:EE_at_5}
\end{figure}

\begin{figure}[ht]
    \centering
    \includegraphics[width=16cm]{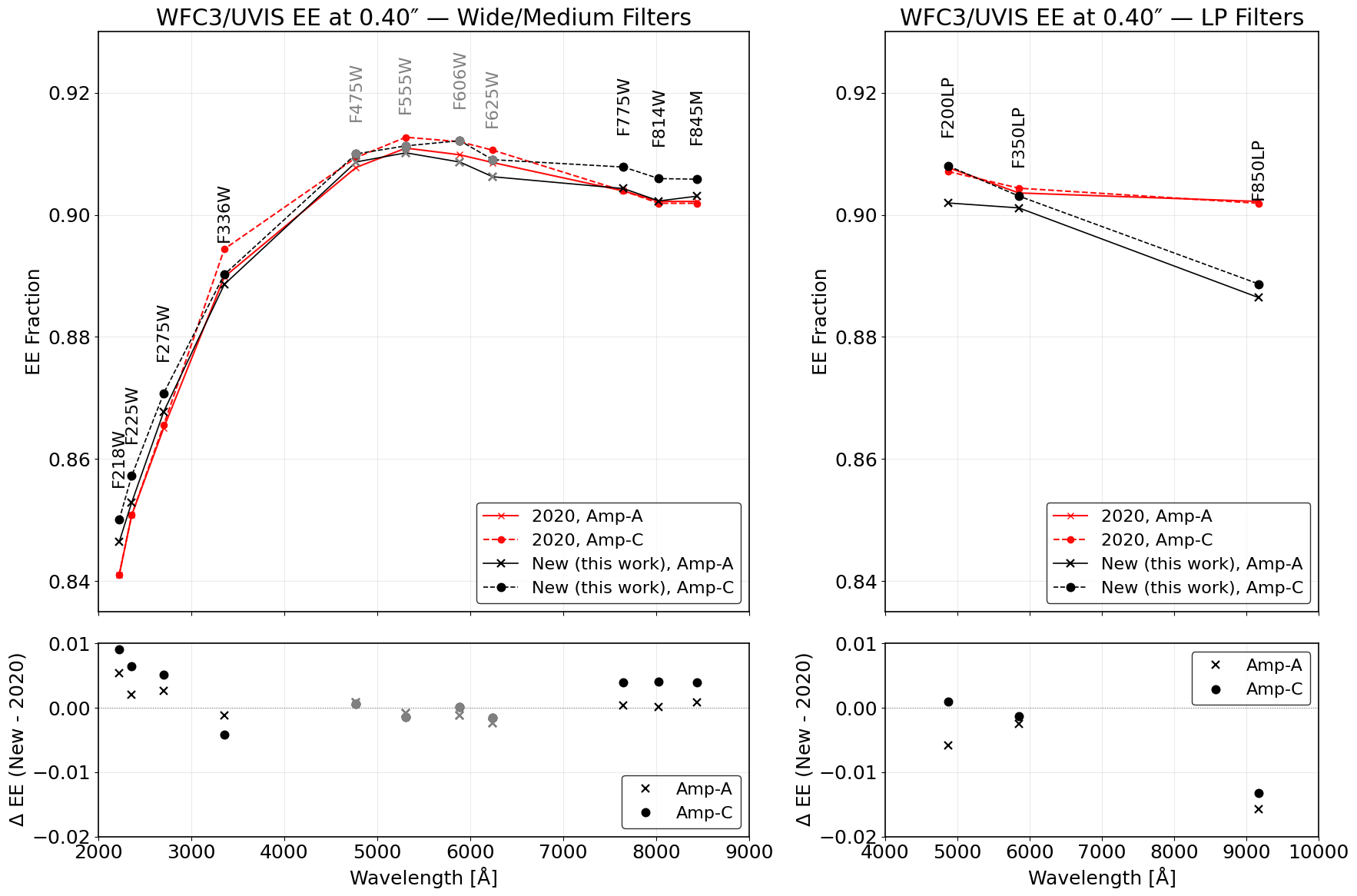}
\caption{(Top panels) Encircled Energy in a 10-pixel aperture radius, EE(10), versus filter pivot wavelength (see Figure~\ref{fig:EE_at_5} for details). (Bottom panels) Difference between the new EE(10) and the 2020 EE(10) values, corresponding to changes required for the UVIS photometric zeropoints. Values are tabulated in Appendix C.
}
    \label{fig:EE_at_10}
\end{figure}

Updated EE tables for UVIS1 and UVIS2 are provided in Appendix A. The new UVIS2 values are systematically larger than their UVIS1 counterparts at radii of 5 pixels (0.2\unit{\arcsecond}) and 10 pixels (0.4\unit{\arcsecond}), across all wavelengths as shown in Figures \ref{fig:EE_at_5} and \ref{fig:EE_at_10}. The UVIS1 EE tables are derived from observations near the corner of amplifier A, where the PSF is elongated by astigmatism. This corner is sensitive to focus, and the EE is typically lower than in the UVIS2 amp C corner. These findings are consistent with the results from focus-diverse PSFs, where the EE in the amp C corner is similar to the mean over the FoV \citep{2025wfc..rept....5H}. The UVIS2 EE tables are therefore recommended for most use cases, and the UVIS1 EE tables are recommended only for targets in the C512A corner subarray. 

\textbf{Blind application of tabulated EEs should be avoided for small apertures (i.e., r $<$ 7 pixels) where the measured photometry (and the EE fraction) is strongly dependent on the telescope focus and orbital breathing}. Instead, aperture corrections should be derived using isolated stars in the individual science frames, when possible, or PSF cutouts at a similar focus level \citep{2025wfc..rept....5H}.

\vspace{1 cm}

\vspace{1 cm}

\subsection{Implications for WFC3/UVIS Zeropoints}

The UVIS zeropoints are derived from aperture photometry within a radius of 10 pixels (0.4\unit{\arcsecond}), corrected for the additional fraction of flux between 0.4\unit{\arcsecond} and 6\unit{\arcsecond} using EE tables. Changes in the EE(10) values indicate that the photometric zeropoints will also need to be updated (see Table \ref{tab:r10_changes}). In this work, we make the fundamental assumption that two quantities are fixed: the total flux of the star at infinity (6\unit{\arcsecond}), as defined by its CALSPEC models, and the count rate at r=10 pixels. Changes in the EE(10) affect the count rate at infinity and will therefore change the inverse sensitivity (PHOTFLAM) values used to compute zeropoints, where PHOTFLAM is defined as the infinite aperture flux of a target (in erg cm\textsuperscript{-2} s\textsuperscript{-1} Å\textsuperscript{-1}) divided by its infinite aperture count-rate (in electrons per second).

Observers using the previous EE tables and zeropoints can manually update their magnitude calculations if performing aperture photometry at any radius other than r=10 pixels. The amount that observers should correct their magnitudes is equal to the amount that the aperture corrections between r and $r_{ref}=10$ change between 2020 and now. More discussion and an example are provided in Appendix B.

\section{Recomendations for Observers}
\label{sec:reccommendations}
Following publication of this report, we plan to deliver updated aperture correction reference files for \texttt{synphot} incorporating the revised encircled energy (EE) measurements presented here. The new files will be named \texttt{wfc3uvis1/2\textunderscore aper\textunderscore 008\textunderscore syn.fits} and are expected later in 2026.

In parallel, improvements to the UVIS time-dependent sensitivity (TDS) correction are being finalized by Pidgeon et al. (2026, WFC3 ISR in preparation). In that work, an updated image photometry reference file (\texttt{IMPHTTAB}, \texttt{*\textunderscore imp.fits}) for \texttt{calwf3} will incorporate not only the revised slopes but also the updated EE(10) values from this work. Together, these improvements will provide a more consistent WFC3/UVIS photometric calibration framework to achieve sub-1\% photometric accuracy (0.01 mag) across both UVIS CCDs.

In the meantime, observers who performed photometry using the 2020 EE and zeropoint values may correct their calibrated magnitudes by applying a differential correction based on the revised EE values presented in this work. An example illustrating the application of the updated EE correction for F275W is provided in Appendix B.

For small apertures, the EE varies on orbital timescales due to telescope ``breathing," which affects the focus. To reduce systematic uncertainties, aperture corrections should be measured directly from isolated stars in the FLC science data whenever possible. For details, see Section~\ref{sec:EE_diff} and \cite{2025wfc..rept....5H}.

\section{Future Work}
\label{sec:Future_work}
Ongoing efforts are focused on improving the empirical measurement of the UVIS EE at large radii and expanding the set of filters with high-fidelity short-stack EE curves. 
In particular, deeper short-stack PSFs, constructed by combining a larger number of individual exposures, are being investigated as a means of empirically measuring the EE beyond 2\unit{\arcsecond}. Such datasets have the potential to reduce reliance on the optical model at large radii and enable a more uniform empirical treatment across filters.

In parallel, we are evaluating the extension of this analysis to filters with EE values derived from 2016 short-stacks, including narrowband filters. Expanding the methodology to include these filters would improve the internal consistency of the WFC3 photometric calibration and further support high-precision UVIS science.

\section{Conclusions}
\label{sec:Conclusions}

In this report, we present updated WFC3/UVIS encircled energy curves for a subset of filters last delivered in 2020. We reanalyzed the same short stack PSFs used in the 2020 delivery (and 2016 short stacks where 2020 data are unavailable). Using improved centroiding and a more consistent normalization strategy, we tie the short stack PSFs to the deep PSFs and provide a robust test of the WFC3 optical model over a broad wavelength range. The new results resolve previously identified discrepancies for small apertures compared to a large study of archival PSFs and will enable a more accurate photometric calibration by improving the EE estimates at 0.4\unit{\arcsecond}.

The key conclusions of this work are summarized below:
\begin{itemize}

\item \textbf{New EE tables are provided in Appendix A. UVIS2 EE tables are recommended for general use; UVIS1 EE tables
are recommended only for targets in the C512A corner subarray.}

\item \textbf{Observers using previous EE values can manually correct their photometry.}
A new photometry reference file will be delivered later in 2026 and will incorporate the new EE(10) values. Until then, observers may follow the example in Appendix B to correct the observed flux for any changes to the EE values. 

\item \textbf{The new EE tables are in close agreement with the results from focus-diverse PSFs for small apertures.}
Refining the centroid locations in the short-stack drizzle-combined PSFs leads to significant changes in the shape of the encircled energy curves at radii smaller than 10 pixels, where the EE is most sensitive to centering errors. The revised EE between r = 5 and 10 pixels yields aperture corrections that agree well with the focus-diverse PSF-based measurements reported by \cite{2025wfc..rept....5H}, resolving systematic discrepancies in the 2020 EE tables for three UVIS filters (F275W, F336W, F814W).

\item \textbf{Our updated methodology primarily affects small aperture radius EE values.} 
 The updated EE values differ from the 2020 delivery by up to $ 0.034$ at r=5 pixels (0.2\unit{\arcsecond}) and $0.016$ at r=10 pixels (0.4\unit{\arcsecond}).
 
\item \textbf{Zeropoint changes are small but significant.}
The combined effects of improved centroiding and revised normalization introduce changes in the photometric zeropoints of up to 0.02 mag. Most re-delivered filters see a change on the order of 0.005 mag for at least one chip. While modest, these changes are relevant for high-precision photometry and will be used to update the UVIS zeropoints later in 2026.

\end{itemize}

\section{Acknowledgments}
 The authors would like to thank WFC3 team reviewers Norman Grogin and Peter R. McCullough for their thorough review and suggested improvements to this report. We thank Aidan Pidgeon for testing the new EE curves and their effect on the UVIS zeropoints, Mariarosa Marinelli for testing the new synphot file, and the WFC3 Photometry Team for many helpful discussions. 
\newpage

\bibliography{ref}
\bibliographystyle{aasjournal}
\nocite{*}

\section{Appendix A }

\begin{minipage}[!h]{1.0\linewidth} \vspace{3ex}
\normalsize
    \centering 
    \def\arraystretch{1.3} 
    \textbf{UVIS2} \\
    \vspace{3mm} 
    \begin{tabular}{|l|c|c|c|c|c|c|c|c|c|c|} \hline
\textbf{Aper} & \textbf{2} & \textbf{3} & \textbf{4} &  \textbf{5} & \textbf{6} & \textbf{7} & \textbf{8} & \textbf{9} & \textbf{10} & \textbf{50} \\ 
\textbf{  } & \textbf{0.08"} & \textbf{0.12"} & \textbf{0.16"} &  \textbf{0.20"} & \textbf{0.24"} & \textbf{0.28"} & \textbf{0.32"} & \textbf{0.36"} & \textbf{0.40"} & \textbf{2.0"} \\ \hline \hline

F218W & 0.542 & 0.686 & 0.752 & 0.787 & 0.807 & 0.821 & 0.832 & 0.841 & 0.850 & 0.987 \\ \hline
F225W & 0.570 & 0.709 & 0.769 & 0.800 & 0.818 & 0.831 & 0.841 & 0.850 & 0.857 & 0.987 \\ \hline
F275W & 0.575 & 0.720 & 0.782 & 0.815 & 0.834 & 0.847 & 0.856 & 0.864 & 0.871 & 0.986 \\ \hline
F336W & 0.592 & 0.740 & 0.802 & 0.835 & 0.854 & 0.867 & 0.877 & 0.884 & 0.890 & 0.986 \\ \hline
F438W* & 0.613 &0.765 & 0.822 & 0.853 & 0.872 &	0.884 &	0.894 & 0.902 &	0.909 & 0.987  \\ \hline
F475W* & 0.593 & 0.755 & 0.819 & 0.851 & 0.871 & 0.884 & 0.894 & 0.902 & 0.909 & 0.985 \\ \hline
F555W* & 0.601 & 0.765 & 0.826 & 0.854 & 0.875 & 0.888 & 0.898 & 0.906 & 0.913 & 0.983 \\ \hline
F606W* & 0.588 & 0.755 & 0.822 & 0.851 & 0.871 & 0.887 & 0.897 & 0.905 & 0.912 & 0.980 \\ \hline
F625W* & 0.579 & 0.747 & 0.818 & 0.846 & 0.867 & 0.884 & 0.896 & 0.904 & 0.911 & 0.979 \\ \hline
F775W & 0.561 & 0.720 & 0.813 & 0.845 & 0.860 & 0.874 & 0.889 & 0.901 & 0.908 & 0.976 \\ \hline
F814W & 0.540 & 0.697 & 0.799 & 0.840 & 0.857 & 0.871 & 0.885 & 0.897 & 0.906 & 0.976 \\ \hline
F845M & 0.533 & 0.684 & 0.794 & 0.839 & 0.856 & 0.869 & 0.881 & 0.895 & 0.906 & 0.975 \\ \hline \hline
F200LP & 0.537 & 0.718 & 0.803 & 0.842 & 0.863 & 0.879 & 0.891 & 0.901 & 0.908 & 0.985 \\ \hline
F350LP & 0.575 & 0.740 & 0.811 & 0.843 & 0.863 & 0.877 & 0.888 & 0.896 & 0.903 & 0.980 \\ \hline
F850LP & 0.496 & 0.636 & 0.749 & 0.811 & 0.834 & 0.848 & 0.861 & 0.875 & 0.889 & 0.975 \\ \hline

    \end{tabular}
    
\captionsetup{type=table}
\captionof{table}{UVIS2 EE for commonly-used filters for aperture radii between 2 and 10 pixels (0.08 - 0.40") and at 50 pixels (2.0"). \textit{Filters marked with an asterix are provided for convenience, but are unchanged from the 2020 values (see text for details)}. For aperture radii $\leq 10$ pixels, the UVIS2 EE is generally larger than in UVIS1 (see Figures 9 and 10). The UVIS2 EE is measured in the corner C512C subarray, where PSF shape is similar to the center of the detector and the mean over the FoV (see Figure 8). \textbf{UVIS2 EE tables are recommended for most use cases} and are therefore listed first in this report. }
    \label{tab:EE_all_UVIS2}
\vspace{2ex}
\end{minipage} 

\begin{minipage}[!h]{1.0\linewidth} \vspace{3ex}
\normalsize
    \centering 
    \textbf{UVIS1} \\
    \vspace{3mm} 
    \def\arraystretch{1.3} 
    \begin{tabular}{|l|c|c|c|c|c|c|c|c|c|c|} \hline
    
\textbf{Aper} & \textbf{2} & \textbf{3} & \textbf{4} &  \textbf{5} & \textbf{6} & \textbf{7} & \textbf{8} & \textbf{9} & \textbf{10} & \textbf{50} \\ 
\textbf{  } & \textbf{0.08"} & \textbf{0.12"} & \textbf{0.16"} &  \textbf{0.20"} & \textbf{0.24"} & \textbf{0.28"} & \textbf{0.32"} & \textbf{0.36"} & \textbf{0.40"} & \textbf{2.0"} \\ \hline \hline

F218W & 0.547 & 0.677 & 0.741 & 0.778 & 0.801 & 0.816 & 0.828 & 0.837 & 0.846 & 0.987 \\ \hline
F225W & 0.540 & 0.681 & 0.748 & 0.786 & 0.809 & 0.824 & 0.836 & 0.845 & 0.853 & 0.987 \\ \hline
F275W & 0.560 & 0.701 & 0.765 & 0.803 & 0.827 & 0.842 & 0.853 & 0.861 & 0.868 & 0.986 \\ \hline
F336W & 0.580 & 0.727 & 0.788 & 0.824 & 0.848 & 0.864 & 0.874 & 0.882 & 0.889 & 0.986 \\ \hline
F475W* & 0.582 & 0.746 & 0.809 & 0.840 & 0.862 & 0.879 & 0.891 & 0.901 & 0.908 & 0.985 \\ \hline
F555W* & 0.575 & 0.745 & 0.812 & 0.842 & 0.864 & 0.881 & 0.893 & 0.903 & 0.911 & 0.983 \\ \hline
F606W* & 0.572 & 0.742 & 0.813 & 0.842 & 0.861 & 0.878 & 0.891 & 0.901 & 0.910 & 0.980 \\ \hline
F625W* & 0.550 & 0.730 & 0.809 & 0.841 & 0.861 & 0.878 & 0.891 & 0.900 & 0.909 & 0.979 \\ \hline
F775W & 0.541 & 0.700 & 0.802 & 0.841 & 0.857 & 0.870 & 0.884 & 0.896 & 0.904 & 0.976 \\ \hline
F814W & 0.527 & 0.686 & 0.792 & 0.836 & 0.854 & 0.868 & 0.881 & 0.893 & 0.902 & 0.976 \\ \hline 
F845M & 0.535 & 0.685 & 0.796 & 0.840 & 0.856 & 0.868 & 0.879 & 0.893 & 0.903 & 0.975 \\ \hline
\hline
F200LP & 0.546 & 0.715 & 0.796 & 0.832 & 0.853 & 0.869 & 0.883 & 0.893 & 0.902 & 0.985 \\ \hline
F350LP & 0.550 & 0.723 & 0.797 & 0.832 & 0.854 & 0.871 & 0.884 & 0.894 & 0.901 & 0.980 \\ \hline
F850LP & 0.485 & 0.628 & 0.745 & 0.809 & 0.834 & 0.848 & 0.861 & 0.873 & 0.886 & 0.975 \\ \hline

    \end{tabular}
    
\captionsetup{type=table}
\captionof{table}{Same as Table \ref{tab:EE_all_UVIS2}, but for UVIS1. The UVIS1 EE is measured within $\sim250$ pixels of the corner of amp A where the EE at radii $\leq 10$ pixels is typically lower than mean EE over the FoV.  \textbf{UVIS1 EE tables are recommended only for targets in the C512A subarray.} }
    \label{tab:EE_all_UVIS1}
\vspace{2ex}
\end{minipage}

\newpage
\section{Appendix B}

If measuring the magnitude using an aperture of radius r=10, EE and ZP changes will cancel each other out, as the larger aperture correction (which is subtracted in magnitude space) is offset by an equal addition to the zeropoint. Thus, observers do not need to correct their magnitude measurements when the measurements are made with an aperture of radius 10 pixels.

However, if using any other aperture ($r$), the change in magnitude is equal to the difference in the aperture correction ($r$ to $r_{ref}=10$) between the two EE deliveries (New and 2020).

\begin{equation}
\centering
\label{eqn:D_ZP_def}
\begin{aligned}
\Delta M = \Delta AP_{10,(New-2020)} = (-2.5log_{10}\frac{EE_{r,New}}{EE_{r=10,New}}) - (-2.5log_{10}\frac{EE_{r,2020}}{EE_{r=10,2020}})
\end{aligned}
\end{equation}

\vspace{0.5 cm}
\subsection*{Example}
As a concrete example, if an observer had performed aperture photometry on a star on \textbf{UVIS2} in the \textbf{F275W} filter using a radius of \textbf{r=5 pixels} and the 2020 EE tables, their final magnitude value would change as follows: 

\begin{equation}
\centering
\label{eqn:D_ZP_applied}
\begin{aligned}
\Delta M = \Delta AP_{10,(New-2020)} = (-2.5log_{10}\frac{0.815 }{0.871}) - (-2.5log_{10}\frac{0.790}{0.866}) = -0.027 mag
\end{aligned}
\end{equation}

where the $EE_{New}$ values are from Appendix A and the $EE_{2020}$ values are from the UVIS Encircled Energy page\footnote{https://www.stsci.edu/hst/instrumentation/wfc3/data-analysis/photometric-calibration/uvis-encircled-energy}). 

\vspace{0.5 cm}
The change in magnitude -0.027 mag is equal to the \textbf{$\Delta$mag} value for F275W UVIS2 in last column of Table \ref{tab:compare_apcorr}. This is also plotted in Figure \ref{fig:phylo_apcorr} with green horizontal lines showing the change in aperture correction for UVIS2 between 5 and 10 pixels.

\newpage
\section{Appendix C}

\begin{minipage}[t]{1.0\linewidth} \vspace{2.5ex}
\normalsize
    \centering
    \def\arraystretch{1.0}
    \begin{tabular}{|l|c|c|c|r|r|} \hline
    
\textbf{Filter}  & \textbf{Amp}  & \specialcell{\textbf{EE(5)} \\ 2020 }& \specialcell{\textbf{EE(5)} \\ 2026} &\specialcell{\textbf{$\Delta EE(5)$} \\ 2026-2020 } & \textbf{$\Delta$mag} \\ \hline \hline
    F218W    & A & 0.766 & 0.778 & +0.012\  & \textbf{-0.017}  \\ 
             & C & 0.766 & 0.787 & +0.021\  & \textbf{-0.029}  \\ \hline
    F225W    & A & 0.780 & 0.786 & +0.006\  &         -0.008   \\ 
             & C & 0.780 & 0.800 & +0.020\  & \textbf{-0.027}  \\ \hline
    F275W    & A & 0.793 & 0.803 & +0.010\  & \textbf{-0.014}  \\ 
             & C & 0.790 & 0.815 & +0.025\  & \textbf{-0.034}   \\ \hline
    F336W    & A & 0.816 & 0.824 & +0.007\  &         -0.009   \\ 
             & C & 0.840 & 0.835 & -0.005\  &         +0.007   \\ \hline \hline
    F475W*   & A & 0.840 & 0.840 & +0.000\  &         -0.000   \\ 
             & C & 0.851 & 0.852 & +0.001\  &         -0.001   \\ \hline
    F555W*   & A & 0.842 & 0.841 & -0.001\  &         +0.002   \\ 
             & C & 0.854 & 0.852 & -0.002\  &         +0.003   \\ \hline
    F606W*   & A & 0.842 & 0.843 & +0.001\  &         -0.001   \\ 
             & C & 0.851 & 0.852 & +0.001\  &         -0.001   \\ \hline
    F625W*   & A & 0.841 & 0.837 & -0.003\  &         +0.004   \\ 
             & C & 0.846 & 0.845 & -0.001\  &         +0.002   \\ \hline
    F775W    & A & 0.842 & 0.841 & -0.001\  &         +0.001   \\ 
             & C & 0.842 & 0.845 & +0.003\  &         -0.004   \\ \hline
    F814W    & A & 0.836 & 0.836 & +0.000\  &         -0.000   \\ 
             & C & 0.821 & 0.840 & +0.019\  & \textbf{-0.024}  \\ \hline
    F845M    & A & 0.836 & 0.840 & +0.004\  &         -0.006   \\ 
             & C & 0.821 & 0.839 & +0.018\  & \textbf{-0.023}   \\ \hline \hline
    F850LP   & A & 0.836 & 0.809 & -0.027\  & \textbf{+0.036}   \\ 
             & C & 0.821 & 0.811 & -0.010\  & \textbf{+0.014}   \\ \hline 
    F200LP   & A & 0.840 & 0.832 & -0.008\  &         +0.010   \\ 
             & C & 0.808 & 0.842 & +0.034\  & \textbf{-0.044}   \\ \hline
    F350LP   & A & 0.834 & 0.832 & -0.003\  &         +0.003   \\ 
             & C & 0.843 & 0.843 & +0.000\  &         -0.000   \\ \hline \hline
    \end{tabular}
    
\captionsetup{type=table}
\captionof{table}{UVIS EE for an aperture radius of 5 pixels (0.2"). Columns list the filter, CCD amplifier, the 2020 EE(5) value, the updated EE(5) value, and the difference relative to the 2020 values. The last column gives the resulting change in magnitude for apertures at r=5 pixels. E.g. for F218W amp C, $\Delta mag$ = -2.5*log10(0.787/0.766) = -0.029 mag. (Filters with $\Delta mag$ values larger than 0.01 mag are shown in bold.) *Short stack PSFs for four VIS filters were tested for completeness. These agree with the 2020 values to within a few milli-mag and will therefore remain unchanged in the 2026 synphot table.
}
    \label{tab:r5_changes}
\vspace{2.5ex}
\end{minipage}

\begin{minipage}[t]{1.0\linewidth} \vspace{2.5ex}
\normalsize
    \centering
    \def\arraystretch{1.0}
    \begin{tabular}{|l|c|c|c|r|r|} \hline
    
\textbf{Filter}  & \textbf{Amp}  & \specialcell{\textbf{EE(10)} \\ 2020 }& \specialcell{\textbf{EE(10)} \\ 2026 } &\specialcell{\textbf{$\Delta EE(10)$} \\ 2026 - 2020 } & \textbf{$\Delta$mag} \\ \hline \hline
    F218W   & A & 0.841 & 0.846 & +0.005\  & \textbf{-0.007}   \\
            & C & 0.841 & 0.850 & +0.009\  & \textbf{-0.012}   \\ \hline
    F225W   & A & 0.851 & 0.853 & +0.002\           &-0.003   \\ 
            & C & 0.851 & 0.857 & +0.006\  & \textbf{-0.008}   \\ \hline
    F275W   & A & 0.865 & 0.868 & +0.003\           &-0.003   \\ 
            & C & 0.866 & 0.871 & +0.005\  & \textbf{-0.006}   \\ \hline
    F336W   & A & 0.890 & 0.889 & -0.001\           &+0.002   \\ 
            & C & 0.894 & 0.890 & -0.004\  & \textbf{+0.005}   \\ \hline \hline
    F475W*  & A & 0.908 & 0.909 & +0.001\           & -0.001   \\ 
            & C & 0.909 & 0.910 & +0.001\           & -0.001   \\ \hline
    F555W*  & A & 0.911 & 0.910 & -0.001\           & +0.001   \\ 
            & C & 0.913 & 0.911 & -0.001\           & +0.002   \\ \hline
    F606W*  & A & 0.910 & 0.909 & -0.001\           & +0.001   \\ 
            & C & 0.912 & 0.912 & +0.000\           & -0.000   \\ \hline
    F625W*  & A & 0.909 & 0.906 & -0.002\           & +0.003   \\ 
            & C & 0.911 & 0.909 & -0.002\           & +0.002   \\ \hline
    F775W   & A & 0.904 & 0.904 & +0.000\           & -0.000   \\ 
            & C & 0.904 & 0.908 & +0.004\   & \textbf{-0.005}   \\ \hline
    F814W   & A & 0.902 & 0.902 & +0.000\           & -0.000   \\ 
            & C & 0.902 & 0.906 & +0.004\   & \textbf{-0.005}   \\ \hline
    F845M   & A & 0.902 & 0.903 & +0.001\           & -0.010   \\ 
            & C & 0.902 & 0.906 & +0.004\   & \textbf{-0.005}   \\ \hline \hline
    
    F850LP  & A & 0.902 & 0.886 & -0.016\   & \textbf{+0.019}   \\ 
            & C & 0.902 & 0.889 & -0.013\   & \textbf{+0.016}   \\ \hline 
    F200LP  & A & 0.908 & 0.902 & -0.006\   & \textbf{+0.007}   \\ 
            & C & 0.907 & 0.908 & +0.001\           & -0.001   \\ \hline
    F350LP  & A & 0.904 & 0.901 & -0.002\           & +0.003   \\ 
            & C & 0.904 & 0.903 & -0.001\           & +0.002   \\ \hline \hline
    \end{tabular}
    
\captionsetup{type=table}
\captionof{table}{Same as Table \ref{tab:r5_changes}, but for an aperture radius of 10 pixels (0.4\unit{\arcsecond}).
The last column reports the change in the photometric zeropoint due to the updated EE(10) value. Filters with $\Delta mag$ values of 0.005 mag or larger are shown in bold, e.g. for F218W amp C, $\Delta mag$= -2.5*log10(0.850/0.841)=-0.012 mag. 
}
    \label{tab:r10_changes}
\vspace{2.5ex}
\end{minipage}

\newpage
\section{Appendix D}

\begin{figure}[h]
    \centering
    \includegraphics[width=16cm]{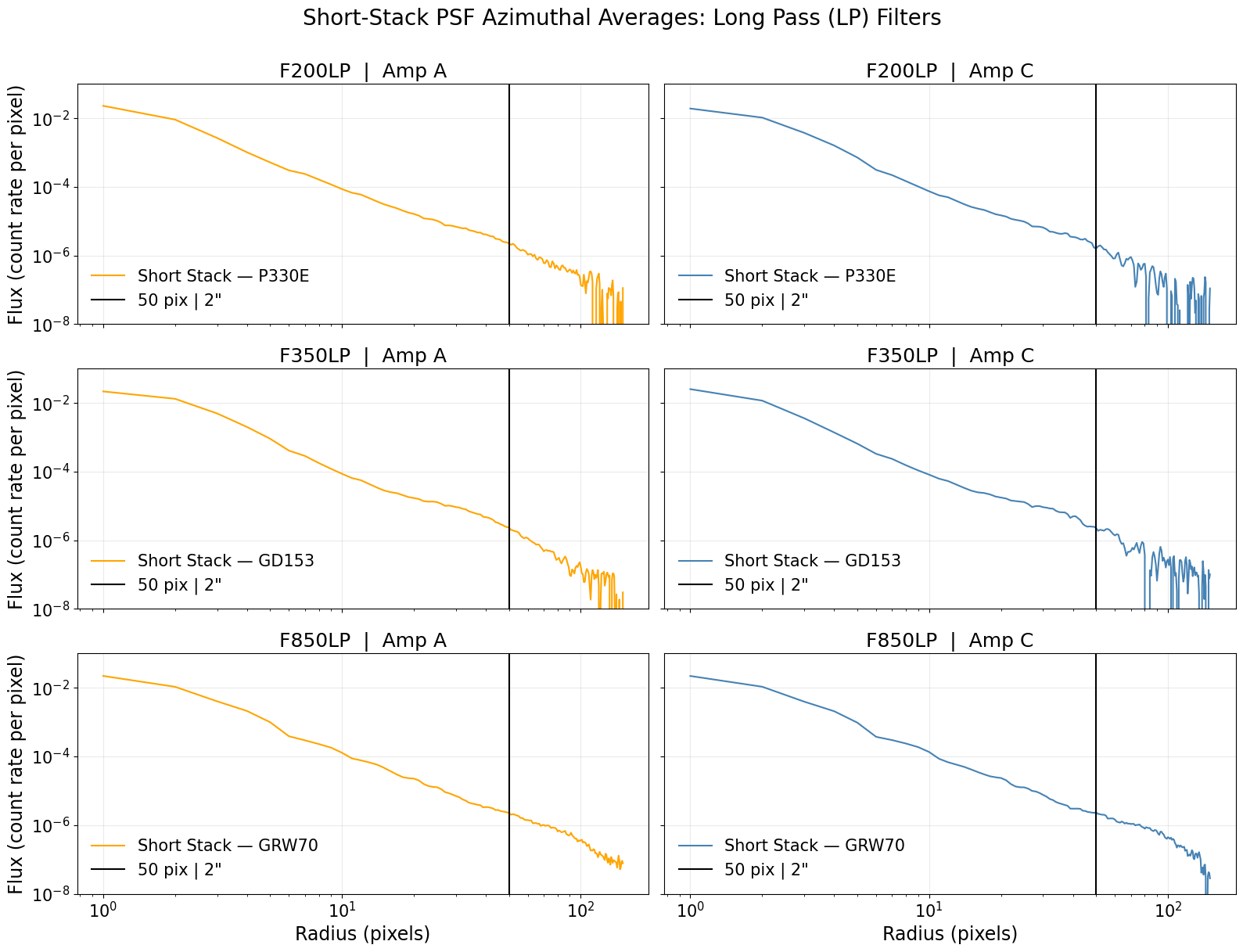}
    \caption{Azimuthal average flux vs radius for short-stack images in LP filters.}
    \label{fig:LP_SN}
\end{figure}

\begin{figure}[ht]
    \centering
    \includegraphics[width=16cm]{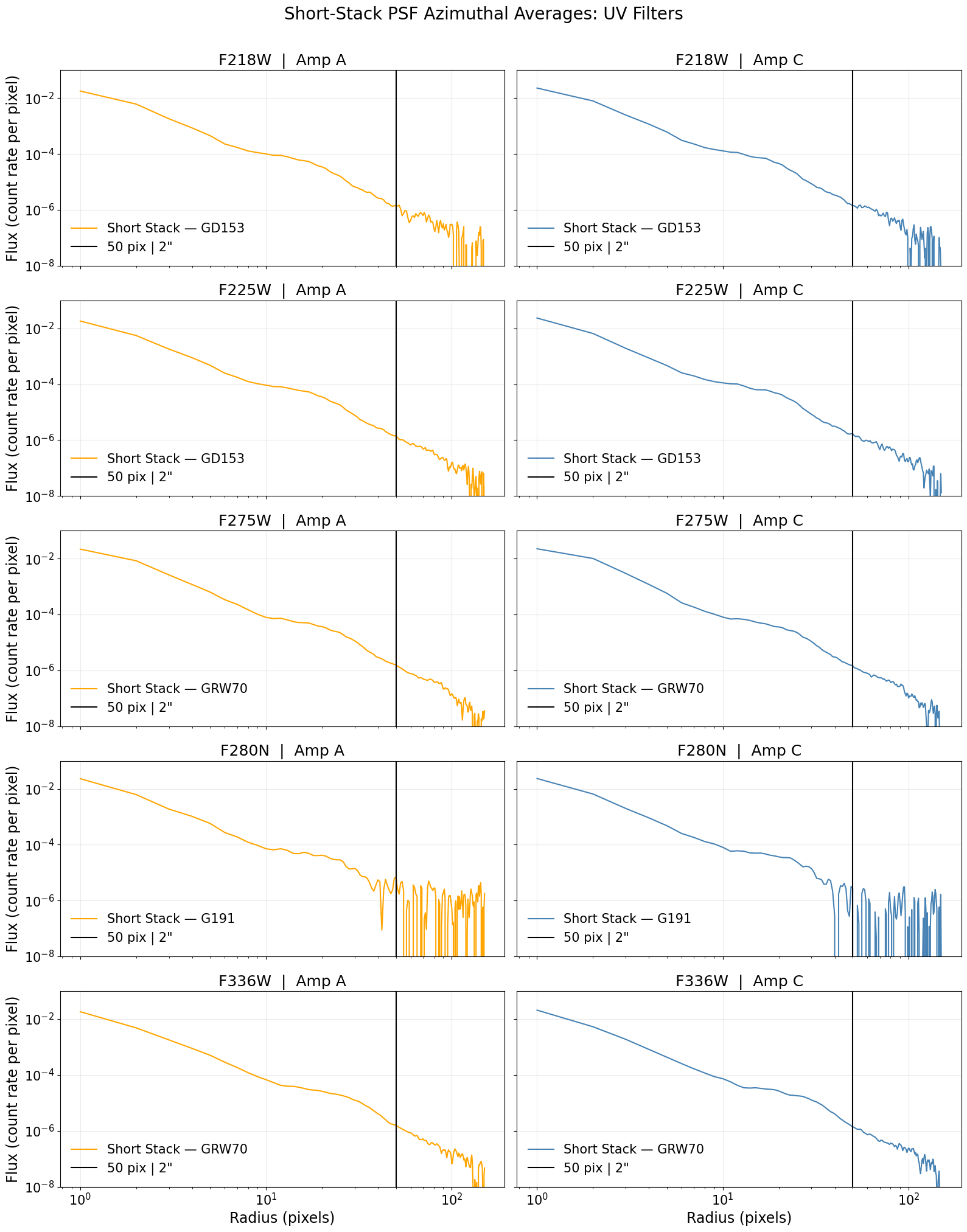}
    \caption{Azimuthal average flux vs radius for short-stack images in UV filters.}
    \label{fig:UV_SN}
\end{figure}

\begin{figure}[ht]
    \centering
    \includegraphics[width=14cm]{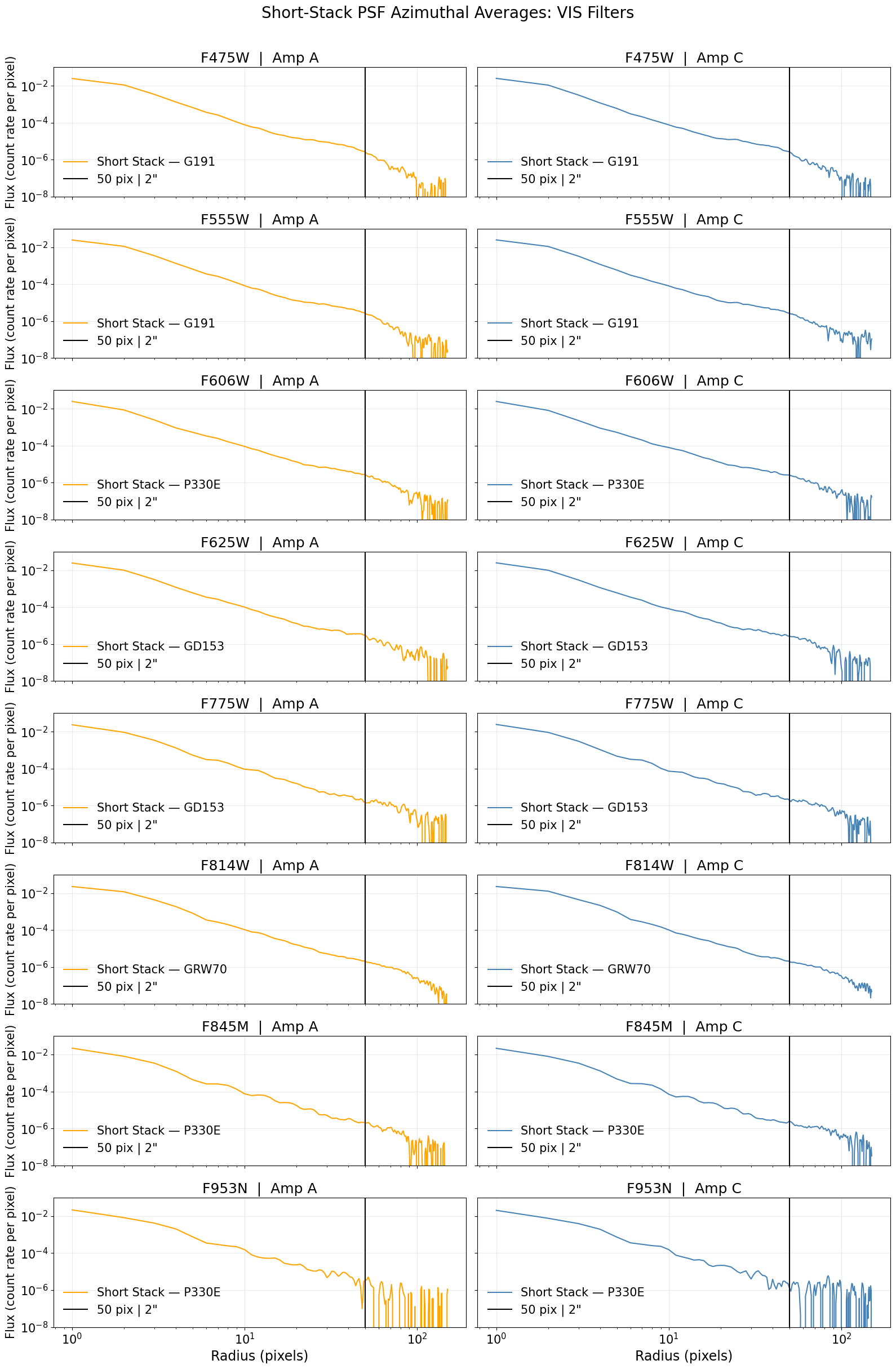}
    \caption{Azimuthal average flux vs radius for short-stack images in VIS filters.}
    \label{fig:VIS_SN}
\end{figure}

\newpage
\section{Appendix E}

\begin{figure}[ht]
    \centering
    \includegraphics[width=12.5cm]{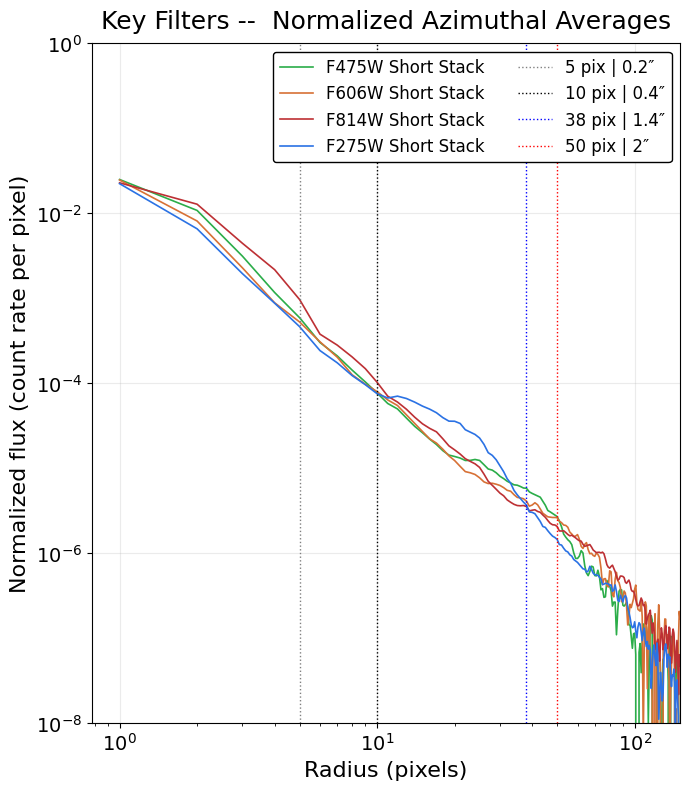}
    \caption{Normalized azimuthal average flux (count rate per pixel) from short-stack PSFs in four key filters across the WFC3/UVIS wavelength range (F275W, F475W, F606W, F814W). The per-pixel fluxes were normalized by the total flux at a radius of 6", as predicted by \href{https://www.stsci.edu/hst/instrumentation/reference-data-for-calibration-and-tools/synphot-throughput-tables}{\texttt{synphot}} using stellar spectra from the CALSPEC database. Redder filters (such as F814W) tend to have more flux in the core of the PSF (radii less than 5 pixels), while bluer filters (such as F275W) have a bump in the azimuthal average between 12 and 35 pixels. 
}
    \label{fig:az_avg_key_filts}
\end{figure}

\end{document}